\documentclass[lettersize,journal]{IEEEtran}
\usepackage{amsmath,amsfonts}
\usepackage{amssymb}
\usepackage[ruled,linesnumbered]{algorithm2e}
\usepackage{caption}
\usepackage{float} 
\usepackage{subcaption}
 \usepackage{cancel}
\usepackage{array}

\usepackage{textcomp}
\usepackage{stfloats}
\usepackage{url}
\usepackage{verbatim}
\usepackage{graphicx}
\usepackage{cite}
\usepackage{bm}
\usepackage{soul}
\usepackage{xcolor}
\usepackage{multirow}
\usepackage{booktabs}
\usepackage{enumerate}
\usepackage{siunitx}
\usepackage{gensymb}
\usepackage{hyperref} 
 \hypersetup{ 
     colorlinks=true, 
     linkcolor=black, 
     filecolor=blue, 
     citecolor = black,       
     urlcolor=blue 
     } 
\usepackage[nolist]{acronym}

\begin{document}
\newacro{DOA}[DOA]{Direction of Arrival}
\newacro{AVSL}[AVSL]{Audio-visual Speaker Localization}
\newacro{TDOA}[TDOA]{Time Difference of Arrival}
\newacro{CE}[CE]{Cross Entropy}
\newacro{FOV}[FOV]{Field-of-View}
\title{Audio Visual Speaker Localization\\ from EgoCentric Views}




\author{
Jinzheng Zhao,
Yong Xu, ~\IEEEmembership{{Senior Member},~IEEE,}
Xinyuan Qian, ~\IEEEmembership{{Senior Member},~IEEE,}

Wenwu Wang, ~\IEEEmembership{{Senior Member},~IEEE,}

\thanks{J. Zhao and W. Wang are with the Centre for Vision, Speech and Signal Processing, University of Surrey, Guildford, GU2 7XH, U.K. E-mail: [j.zhao, w.wang]@surrey.ac.uk.}

\thanks{Y. Xu is with Tencent AI Lab, Bellevue, WA 98004, USA. E-mail: yong.xu.ustc@gmail.com.}

\thanks{X. Qian is with the Department of Computer Science and Technology, University of Science and Technology Beijing, Beijing 100083, China. E-mail: qianxy@ustb.edu.cn.}

\thanks{This research is supported by Tencent AI Lab Rhino-Bird Gift Fund and University of Surrey. This research is also supported by UKRI EPSRC and BBC Prosperity Partnership AI4ME. For the purpose of open access, the authors have applied a Creative Commons Attribution (CC BY) licence to any Author Accepted Manuscript version arising. }
}



\maketitle

\begin{abstract}
The use of audio and visual modality for speaker localization has been well studied in the literature by exploiting their complementary characteristics. However, most previous works employ the setting of static sensors mounted at fixed positions. Unlike them, in this work, we explore the ego-centric setting, where the heterogeneous sensors are embodied and could be moving with a human to facilitate speaker localization. Compared to the static scenario, the ego-centric setting is more realistic for smart-home applications e.g., a service robot. However, this also brings new challenges such as blurred images, frequent speaker disappearance from the field of view of the wearer, and occlusions. In this paper, we study egocentric audio-visual speaker \ac{DOA} estimation and deal with the challenges mentioned above. Specifically, we propose a transformer-based audio-visual fusion method to estimate the relative DOA of the speaker to the wearer, and design a training strategy to mitigate the problem of the speaker disappearing from the camera's view. We also develop a new dataset for simulating the out-of-view scenarios, by creating a scene with a camera wearer walking around while a speaker is moving at the same time. The experimental results show that our proposed method offers promising performance in this new dataset in terms of tracking accuracy. Finally, we adapt the proposed method for the multi-speaker scenario. Experiments on EasyCom show the effectiveness of the proposed model for multiple speakers in real scenarios, which achieves state-of-the-art results in the sphere active speaker detection task and the wearer activity prediction task. The simulated dataset and related code are available at \url{https://github.com/KawhiZhao/Egocentric-Audio-Visual-Speaker-Localization}.

\end{abstract}


\section{Introduction}
\label{introduction}
\IEEEPARstart{V}{ision} and hearing are two important modalities for humans to perceive the world. 
Specifically, vision provides an informative signal, from which we can localize the objects when they are visible. Hearing can assist the visual modality to improve localization robustness. The two modalities can serve as complementary signals of each other for joint localization. For instance, if the audio signal is contaminated by background noise and reverberation, or objects are silent, the visual signal can be used. On the contrary, if objects move outside the camera's \ac{FOV} or being occluded, audio can be used. 

The aim of \ac{AVSL} is to find the positions (either in 3D Cartesian coordinates or \ac{DOA}) of the speakers using both audio and visual modalities, captured by microphones (or microphone arrays) and cameras, respectively. The \ac{AVSL} systems could be used for speaker monitoring \cite{hampapur2005smart}, speech enhancement \cite{loizou2007speech} and speech separation \cite{ong2022audio}.
For the visual modality, off-the-shell advanced algorithms, such as object detectors \cite{redmon2018yolov3}, head detectors \cite{vu2015context} and face detectors \cite{li2019dsfd}, can be used to localize the speakers on the image plane. Traditional methods like color histogram \cite{birchfield2005spatiograms} find the target position by comparing the similarity between the reference image and sub-regions of the whole image. It is often used as complementary measurements if the face detector fails e.g., when the speakers are not facing towards the camera or when the illumination changes~\cite{qian2019multi}. 
For the audio modality, sound source localization algorithms, either parametric- or learning-based, can be used to estimate the speaker's position. 
A popular example of the parametric methods is the \ac{TDOA}-based methods \cite{knapp1976generalized, brutti2008localization, d2012person, zotkin2001multimodal}, which estimate the time delay by maximizing the \ac{TDOA} likelihood. The \ac{TDOA}-based methods are computationally efficient but with poor generalization ability across different acoustic scenarios \cite{brutti2008localization}. The learning-based methods have gained significant attention in recent years, primarily due to the widespread adoption of deep learning techniques and the accessibility of GPU resources. Different from the parametric-based methods which rely on statistical modeling, learning-based methods \cite{he2018deep,he2016deep,wang2021gcc,cao2019polyphonic} find the direct mapping between the input audio features and the speaker locations through training with large-scale annotated data. Thus, they are well generalized to sounds with room reverberation and background noise, and also act as the focus of our paper.
  

Tasks in the egocentric scenarios such as egocentric audio visual diarization \cite{xu2022ava}, forecasting \cite{liu2020forecasting}, and action recognition \cite{kazakos2019epic}, have drawn increasing attention in recent years. 
In contrast to the conventional setting where sensors are stationary, the egocentric scenario presents unique challenges characterized by dynamic camera views and distortions resulting from the wearer's head and body movements. To address these emerging challenges, several egocentric datasets have been introduced to foster advancements in this research domain, notable examples include Ego4D \cite{grauman2022ego4d}, Epic Kitchen \cite{damen2018scaling} and EgoCOM \cite{northcutt2020egocom}.
The task of EgoTracks in Ego4D is localizing the speakers in the image plane. In this case, when a speaker disappears from the view, the track is lost. In our task, we calculate the relative DOA\footnote{We calculate \ac{DOA} in the world coordinates instead of in the image plane.} of the speaker with respect to the wearer, irrespective of whether the speaker is visually observable or not.
To this end, we propose a transformer-based method and considering the character of audio and visual modality that the visual modality is only useful within the \ac{FOV} and audio modality is useful on the sphere space around the wearer, we design a training strategy which separates the contributions from vision and audio. 

To validate the effectiveness of our proposals, we propose a new simulated dataset Egocentric Audio Visual Speaker Localization (Ego-AVSL) for AVSL, which is built upon the Unity game engine.  Specifically, the scene includes two people moving freely in an indoor environment, with one being the speaker and the other wearing a camera and a microphone array.
Examples of this scenario can be found in Fig. \ref{global}. The main reason why we create a simulated dataset is that it can be easily extended to a large quantity without expensive manual annotation. Another reason is because of the privacy issue \cite{fabbri2021motsynth}. Some laws related to information security, such as the General Data Protection Regulation (GDPR) have been established in Europe, claiming that people's visual image data cannot be used without authentication. In addition, experiments in \cite{fabbri2021motsynth} show that the use of additional simulated data for training can improve the performance of the model on real datasets.

The most related works to ours are \cite{donley2021easycom,jiang2022egocentric} where the EasyCom dataset was created with multiple people sitting around a table, talking in turn and ordering meals. The wearer is equipped with a camera to record videos and Google AR glasses to obtain multi-channel audio. However, in this dataset, the subjects are mostly static where the view of the wearer remains unchanged to a large extent. Thus, the subjects are mostly visible in the wearer's view, which is not always the case in egocentric speaker tracking. Even though, EasyCom still stands for a real scenario which can be used for evaluation. 

Our contributions are summarized as follows:
\begin{enumerate}
    
    \item We propose an audio-visual fusion model for egocentric \ac{DOA} estimation and a training strategy which separately processes the audio-only and audio-visual data to tackle challenging scenarios e.g., the speaker moves outside the camera's \ac{FOV}. 
    \item 
    We develop a new dataset, namely Ego-AVSL, to simulate the out-of-view scenarios and facilitate the study of egocentric AVSL.  
    \item Extensive experiments are conducted where the results demonstrate that the audio modality can serve as a complementary signal when the visual modality is missing. Using visual modality can improve the model performance when the speaker is visible.
    \item Further experimental results on the EasyCom dataset show the effectiveness of our model for multiple speaker localization in real scenarios and the pretrained model on the simulated dataset can provide a good initialization.
\end{enumerate}


The remainder of the paper is organized as follows. Section \ref{rw} reviews the related works and Section \ref{formulation} defines the problem and analyses the characters of ego-centric scenarios. Section \ref{methods} describes the architecture of our proposed model. In Section \ref{Dataset}, we introduce the generation process of the proposed dataset and highlight its distinctive features compared to existing datasets in the literature. Section \ref{experiments} shows experimental results and the corresponding visualizations. Section \ref{extension_multiple} demonstrates the model's performance in real scenarios when multiple speakers exist. The last section concludes the paper. 

\section{Related work}
\label{rw}

In this chapter, we summarize related works and highlight their differences from ours.

\subsection{Ego-centric Study}
Ego-centric study has thrived in the past few years as it describes human activities from a first-person perspective, which mimics the way how humans perceive the world. EPIC-KITCHENS \cite{damen2018scaling} is a large-scale dataset that collects first-person video recordings of kitchen activities, which can be used for ego-centric segmentation, object tracking, action recognition, and detection. Ego4D \cite{grauman2022ego4d} is the biggest ego-centric dataset till now, containing over 3,000 hours recorded by participants around the world. This dataset supports the exploration of episodic memory, social understanding, and forecasting. 
Unlike the previous works, we focus on the egocentric AVSL, which explores the complementary characteristics of audio and visual data for \ac{DOA} estimation. 
Egocentric speaker localization plays a pivotal role in audio-visual navigation tasks, which employ an agent to locate a stationary sound source that emits continuous sounds.

\subsection{Sound Source Localization and Speaker Localization}
The challenges of Detection and Classification of Acoustic Scenes and Events (DCASE) and Learning 3D Audio Source (L3DAS) have drawn much attention in sound source localization where an increasing number of learning-based methods \cite{grumiaux2022survey} have recently appeared. In \cite{he2018deep}, a neural network based on MLP and CNN has been proposed to determine DOA. In \cite{vera2018towards}, an end-to-end neural network has been designed, which takes raw waveform as input to predict the 3D coordinates. CRNN is introduced in \cite{adavanne2018sound} to predict the sound event and positions concurrently. A two-stage strategy is proposed in \cite{cao2019polyphonic}, where the sound event is predicted first to assist the location estimation. Instead of just using audio modality, in the task of sound source localization, audio and visual modalities can jointly work to infer the positions of sounding objects. As indicated in \cite{michelsanti2021overview}, fusion methods can be divided into early, intermediate, and late strategies. In particular, early fusion directly combines the extracted features. Intermediate fusion integrates the features of the two modalities and allows multi-modality interaction. Late fusion integrates the two modalities at the decision level.

Speaker localization can be seen as a subtask of sound source localization.  In \cite{qian2021multi}, the early concatenation of GCC-PHAT and the simulated visual features are input to MLP to obtain DOA. A similar idea is also applied in \cite{wu2023multi}, which uses early fusion to integrate the STFT and visual Gaussian distribution. In \cite{afouras2020self}, the audio-visual representation is learned through audio-visual correspondence and contrastive learning. Then the learned embedding can be used for downstream tasks such as audio visual object detection and speech separation. In \cite{wissing2021data}, the audio visual modalities are intermediately fused with dynamic weights, indicating the changing importance of the two modalities at different time steps. In \cite{qian2022deep}, audio and visual streams are fused intermediately. They are input to a variational autoencoder to generate correlation functions. Then beamforming is used to generate the acoustic map. In \cite{qian2022audio}, a new dataset for audio-visual speaker \ac{DOA} estimation is proposed. And a cross-modal attention mechanism is proposed for intermediate fusion. Intermediate fusion is also used for our method as it can use the complementary information of audio and visual modalities.

In this paper, we focus on egocentric speaker localization, which leverages sound source localization and speaker localization methods with audio-visual fusion, and try to mitigate the problems caused by egocentric scenarios such as frequent speaker disappearance. 

\section{Problem Formulation}
\label{formulation}

\subsection{Problem Formulation}
\begin{figure}[tbp]
    \centering
    \includegraphics[width=0.65\columnwidth]{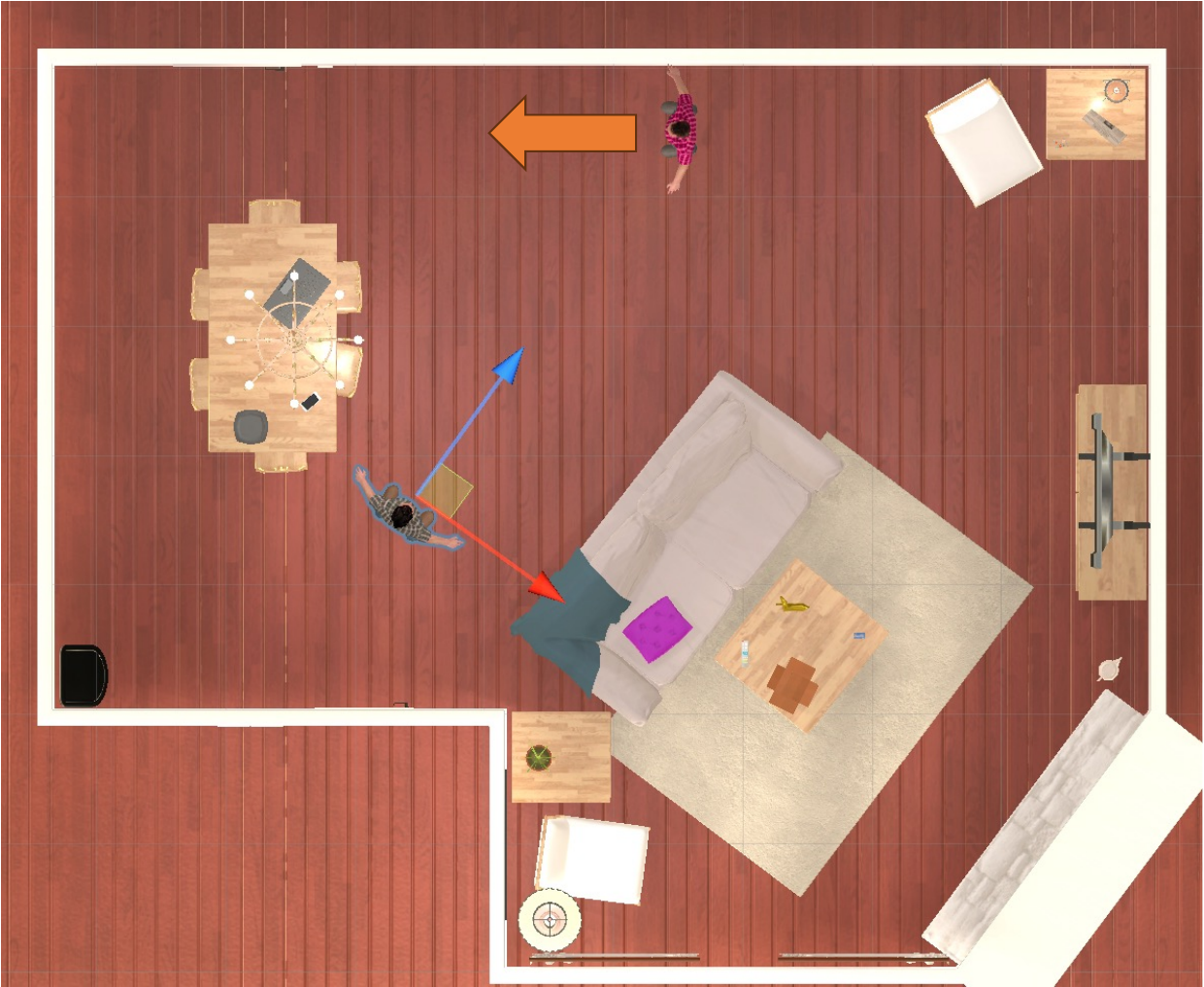}
    \caption{The scenarios of ego-centric speaker localization. The blue arrow of the wearers is the positive north direction and the red arrow is the positive east direction. The blue arrow is also the velocity direction of the wearer. The orange arrow is the velocity direction of the speaker.}
    \label{global}
\end{figure}

As an example, a speaker and a wearer are in the room in their own orientations and positions, and walk randomly. Given an audio clip and a video frame, the purpose is to calculate the relative \ac{DOA} of the speaker of the wearer. This scenario can be seen in Fig. \ref{global}. The positive direction is set as the orientation of the wearer (the blue arrows in Fig. \ref{global}). The coordinates of the speaker are converted into the coordinate system centered on the wearer's position. The ground truth \ac{DOA} is calculated by the tangent value of the converted coordinates of the speaker. Given the position of the speaker $(x_{1}, y_{1}, z_{1}, r_{1})$ and the wearer $(x_{2}, y_{2}, z_{2}, r_{2})$. The relative \ac{DOA} is calculated as:
\begin{equation}
        \theta=\arctan \frac{(x_{2} \sin{r_2} + z_{2} \cos {r_2}) - (x_{1} \sin{r_2} + z_{1} \cos {r_2})} {(x_{2} \cos{r_2} - z_{2} \sin {r_2}) - (x_{1} \cos{r_2} - z_{1} \sin {r_2})}
\end{equation}
 where $\theta$ denotes the azimuth, $(x, y, z)$ is the position and $r$ is the rotation angle, which can be obtained by the quaternion coordinates.

\begin{figure}[tbp]
    \centering
    \includegraphics[width=1\columnwidth]{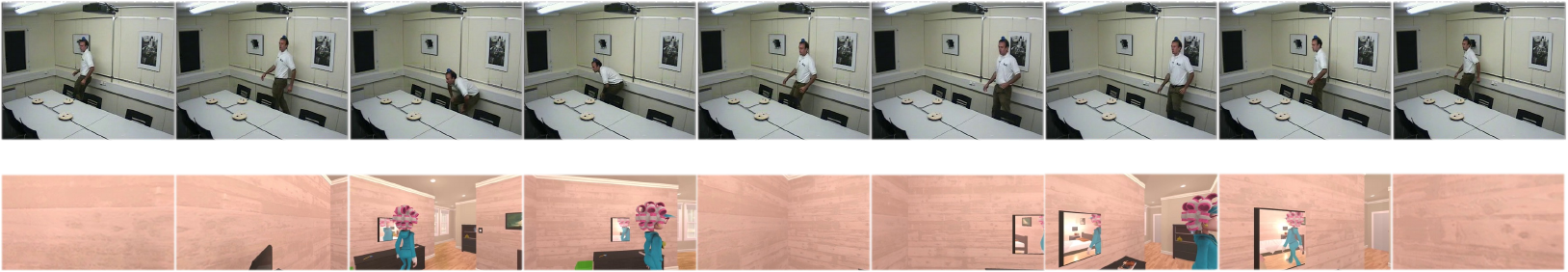}
    \caption{Snapshots of AV16.3 dataset (the up row) and our simulated Ego-AVSL dataset (the bottom row) with the interval of one second.}
    \label{sequence}
\end{figure}

\begin{figure*}
\centering
\begin{minipage}{0.32\textwidth}
  \centering
  \includegraphics[width=\linewidth]{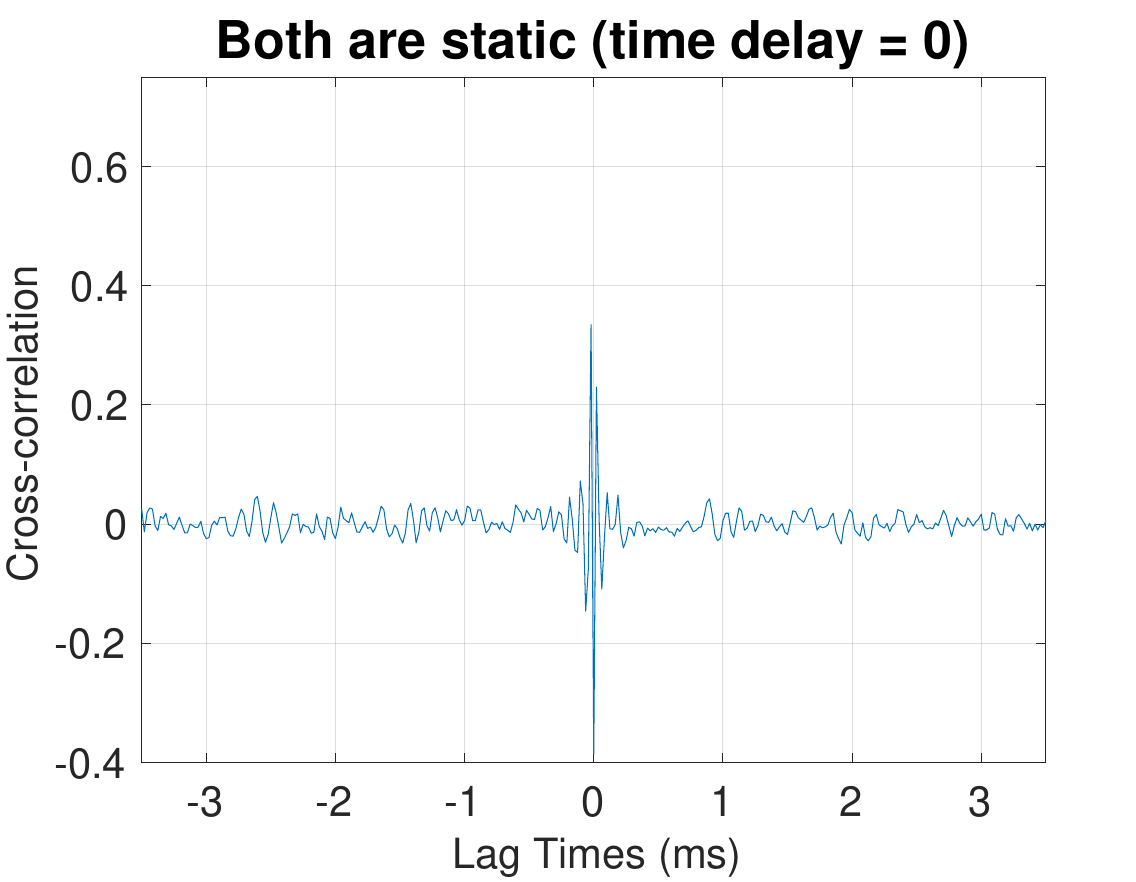} 
  \caption{Time lag estimation by GCC-PHAT when both the wearer and the speaker are static.}
  \label{0moving}
\end{minipage}\hfill
\begin{minipage}{0.32\textwidth}
  \centering
  \includegraphics[width=\linewidth]{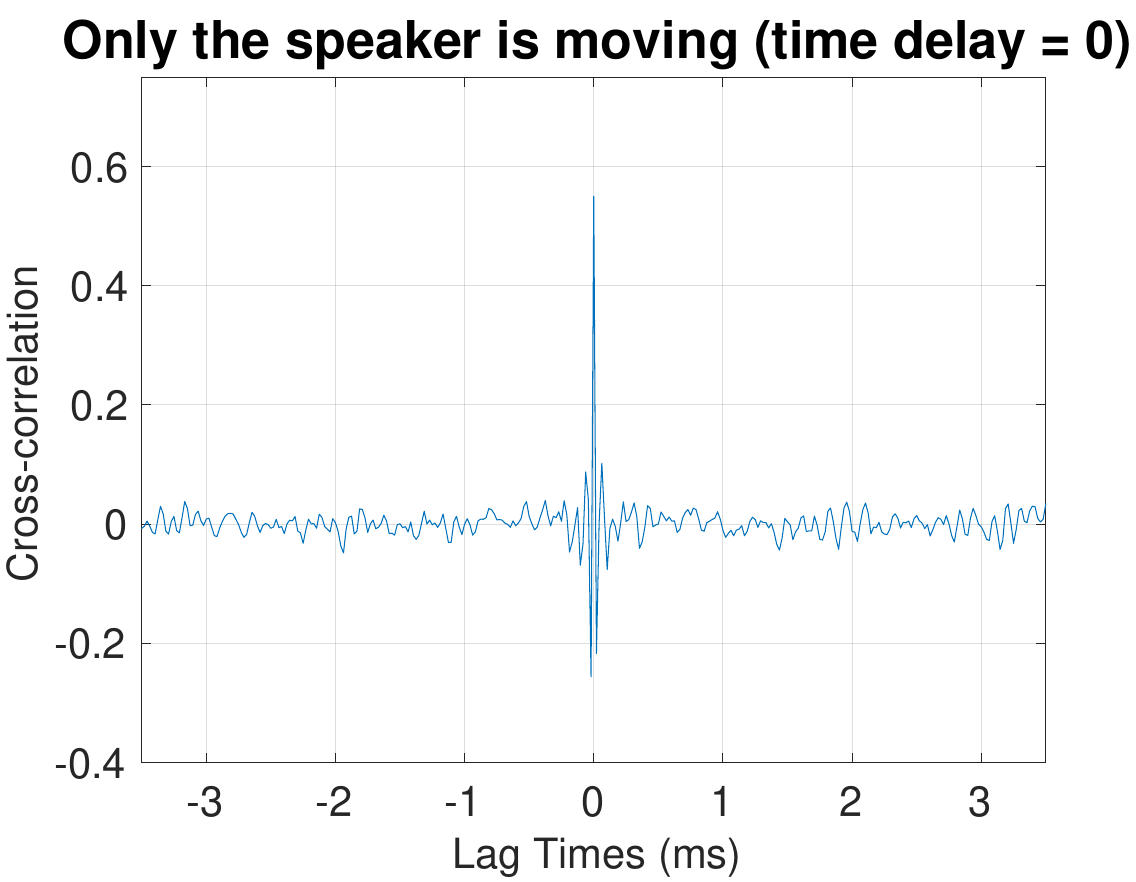}
  \caption{Time lag estimation by GCC-PHAT when the speaker is moving and the wearer is static.}
  \label{1moving}
\end{minipage}\hfill   
\begin{minipage}{0.32\textwidth}
  \centering
  \includegraphics[width=\linewidth]{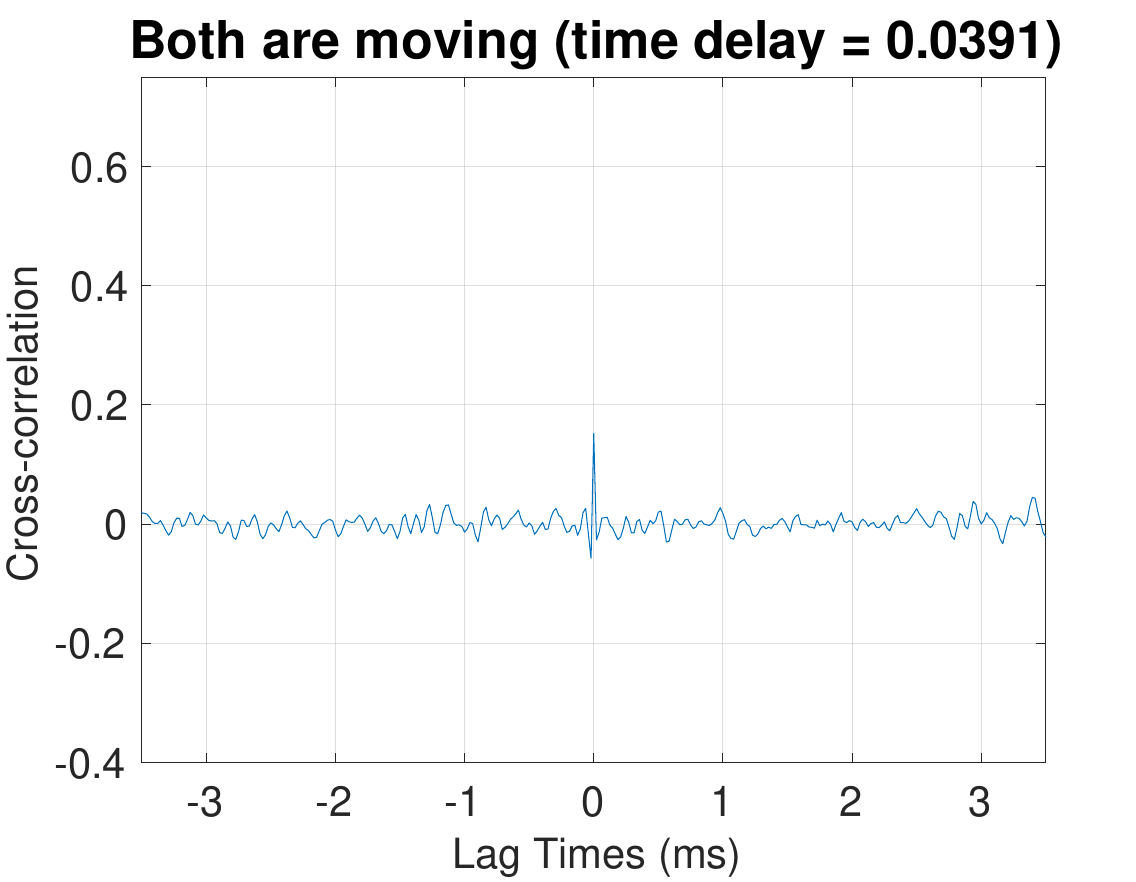}
  \caption{Time lag estimation by GCC-PHAT when both the wearer and the speaker are moving.}
  \label{2moving}
\end{minipage}

\end{figure*}

\subsection{Difference between Egocentric Scenario and General Scenario of Audio Visual Speaker Tracking}
\label{difference}
We discuss the difference and the challenges in terms of audio modality and visual modality, respectively.
\subsubsection{Audio Modality}
In ego-centric scenario, the wearer is also moving, which has a Doppler effect on the received audio and will affect the frequency of the received audio. The change of the frequency is calculated as follows:
\begin{equation}
    \hat{f} = f  \frac{c + v_{r}}{c - v_{s}}
\end{equation}
\noindent where $\hat{f}$ is the frequency of the received signal and $f$ is the frequency of the emitted signal. $v_{r}$ is the receiver's velocity and $v_{s}$ is the source's velocity. $c$ is the sound speed.

In the general speaker tracking scenario, only $v_{s}$ affect the change of frequency. But in the egocentric speaker tracking scenario, both $v_{r}$ and $v_{s}$ contribute to the Doppler effect. 

To validate the difference of Doppler effects, we show the time lag estimation of GCC-PHAT in different scenarios. As the change of frequency will affect the phase of the received audio and GCC-PHAT relies on the phase difference to estimate the time lag. The simulated scenario is in Fig. \ref{global}. People with gray clothes is the wearer and people with red clothes is the speaker. The wearer is equipped with a camera and a microphone array with two microphones, and is facing to the speaker. The blue arrow denotes the velocity direction of the wearer and the orange arrow denotes the velocity direction of the speaker. Three scenarios are simulated:
\begin{itemize}
    \item Both the wearer and the speaker are static.
    \item Only the speaker is moving with a constant velocity.
    \item Both the wearer and the speaker are moving with constant velocities.
\end{itemize}

We record the received audio when the speaker and the wearer are at the same preset points. Although the audios are recorded at the same points among the above scenarios, the difference is that in the second scenario, the speaker has a velocity. And in the last scenario, both the wearer and the speaker have velocities. We ensure that the quaternion of the wearer and the speaker are the same in the three scenarios to remove the influence of the facing direction. We use the audio whose length is equivalent to two visual frames and assume that within such a short period, the wear and the speaker are static.

The simulation results for the three scenarios are shown in Fig. \ref{0moving}, Fig. \ref{1moving} and Fig. \ref{2moving}, respectively. The simulation of the first scenario provides the ideal time lag estimation as there is no Doppler effect. The estimated time delay is 0 as the wearer is facing the speaker and there are no delays for two microphones to receive the audio. If the speaker is moving, the frequency (magnitude and phase) of the received audio is changed, which will affect the time lag estimated by GCC-PHAT as GCC-PHAT relies on the phase relationship of the received binaural audio. The estimated time lag is the same as that in the first scenario. When the wearer is moving as well, the estimation (0.0391) deviates from 0, which shows that the Doppler effect is stronger and affects the time lag estimations. 

If the receivers move faster, like drones \cite{essaadali2016new} and underwater vehicles \cite{jia2020localization}, the larger speeds will have a more obvious Doppler effect.

\subsubsection{Visual Modality}
\label{difference visual}
The difference from general tracking scenarios in visual modality is more straightforward, which is the more frequent speaker disappearance.
We show the sequences of the AV16.3 dataset \cite{lathoud2004av16} and our simulated Ego-AVSL dataset in Fig. \ref{sequence}. In the AV16.3 dataset, the camera is fixed and the speaker is in the view in most cases. While in the Ego-AVSL dataset, due to the movement of the wearer, the speaker is out of view at some moments, which may pose a challenge to data fusion. In \cite{qian2021multi}, the early fusion is used and the features of the audio and visual modality are flattened and concatenated. However, this kind of fusion may degrade the model performance, as the visual image does not contain the speaker's position information when the speaker disappears. 

In this paper, we mainly focus on mitigating the problem in visual modality. Although the Doppler effect is stronger than that in general tracking scenarios, it is still negligible given the low speed of movement. However, the out-of-view problems in visual modality pose challenges in audio-visual fusion.


\section{Proposed Method}
\label{methods}
In this part, we start by introducing the model which can mitigate the out-of-view problem. Then we introduce the EMD loss which considers the continuous ordinal relationships between DOA angles and is proven to be superior to \ac{CE} loss. Finally, we discuss the audio and visual features extracted to represent the spatial information.
\subsection{Model}
\begin{figure}[tbp]
    \centering
    \includegraphics[width=\columnwidth]{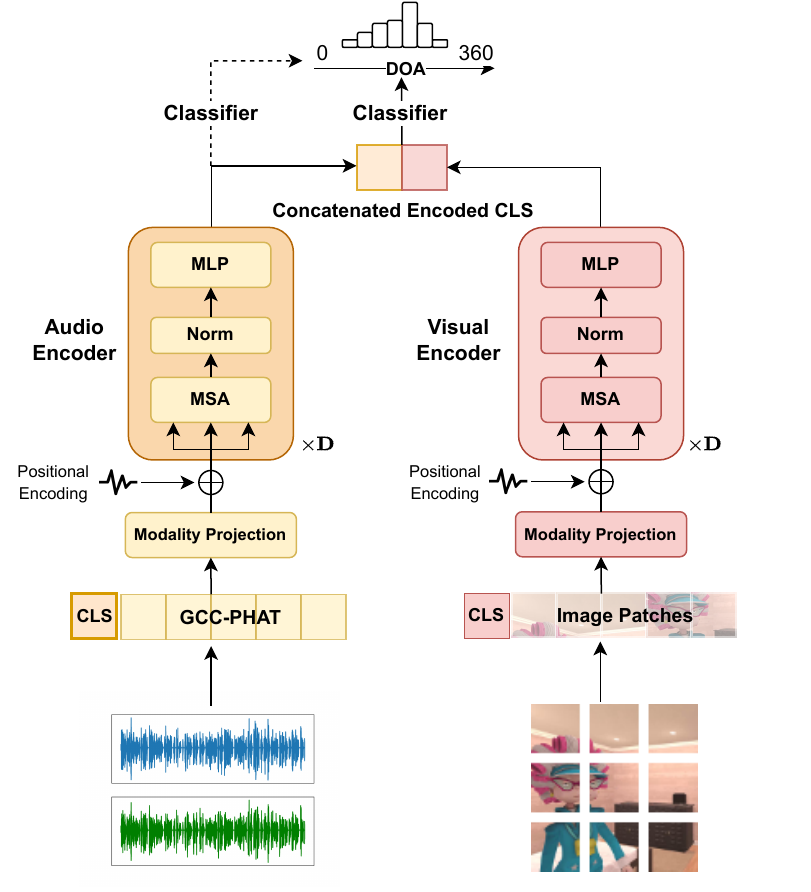}
    \caption{The architecture of the proposed model. The input of audio encoder is binaural waveforms while the input of visual encoder is a sequence of flattened patches. Features from audio and visual modalities are extracted and fused intermediately.}
    \label{model}
\end{figure}

As discussed in Section \ref{difference visual}, in the egocentric scenario, the speaker moves out of the camera view from time to time. Early fusion of audio and visual features will bring redundant visual information when the speaker is not in the view. Besides, it is not robust against temporal misalignment and outliers \cite{wu2019dual}. Late fusion will make the prediction unreliable if the quality of one modality is bad. To this end, we use intermediate fusion with separate training to integrate audio and visual features, aiming to extract useful information and discard redundant parts.
The architecture of the proposed model is shown in Fig. \ref{model}. The audio and visual features go through different encoders and are fused through a cross-attention module. 
We use Transformer encoder as the backbone. 
A $[CLS]_{a}$ token is added at the beginning of the audio feature and another $[CLS]_{v}$ token is added at the beginning of visual feature, aiming to learn the localization information using audio modality and visual modality, respectively. The features are added with the position encoding to retain positional information. The output tokens from encoders are concatenated and go through the cross-attention module, which integrates information from audio and visual modalities. The output of the cross-attention module is concatenated as fused feature and a classifier is used to obtain the \ac{DOA} posterior based on the fused audio-visual feature. The \ac{DOA} estimation process is as follows:
\begin{equation}
    \mathbf{z}^{0}_{a} = [[CLS]_{a}; \mathbf{a}_{1}\mathbf{\textit{A}}; ...; \mathbf{a}_{L_{a}}\mathbf{\textit{A}} ] + \mathbf{a}^{pos}
    \label{eq_audio}
\end{equation}
\begin{equation}
    \mathbf{z}^{0}_{v} = [[CLS]_{v}; \mathbf{v}_{1}\mathbf{\textit{V}}; ...; \mathbf{v}_{L_{v}}\mathbf{\textit{V}} ] + \mathbf{v}^{pos}
    \label{eq_visual}
\end{equation}
\begin{equation}
\hat{\mathbf{z}}^d_{m}=\operatorname{LN}\left(\operatorname{MSA}\left(\mathbf{z}^{d-1}_{m}\right)\right)+\mathbf{z}^{d-1}_{m}, \quad d=1 \ldots D, \quad m = a, v
\label{eq_self}
\end{equation}
\begin{equation}
\mathbf{z}^d_{m}=\operatorname{LN}\left(\operatorname{MLP}\left(\hat{\mathbf{z}}^d_{m}\right)\right)+\hat{\mathbf{z}}^d_{m}, \quad d=1 \ldots D, \quad m = a, v
\label{eq_mlp}
\end{equation}
\begin{equation}
    \hat{p} = \operatorname{MLP}\left(\operatorname{MSA}\left(\mathbf{z}_{a}^{D} \oplus \mathbf{z}_{v}^{D}\right)\right)
    \label{eq_out}
\end{equation}

\noindent where Eq. (\ref{eq_audio}) and Eq. (\ref{eq_visual}) are the input representations of audio and visual modalities. Eq. (\ref{eq_self}) and Eq. (\ref{eq_mlp}) denote the multi-head self-attention and feed-forward processes. Eq. (\ref{eq_out}) predicts the \ac{DOA} given the joint audio visual features. $\mathbf{a}_{1},..., \mathbf{a}_{L_{a}}$ denotes the audio feature with length $L_{a}$ and $\mathbf{v}_{1},..., \mathbf{v}_{L_{v}}$ denotes the visual feature with length $L_{v}$. $\mathbf{\textit{A}}$ and $\mathbf{\textit{V}}$ denote the audio and visual embedding layers, respectively, which project the audio features and visual features into the same dimension. $\mathbf{a}^{pos}$ and $\mathbf{v}^{pos}$ represent the position encoding for audio and visual modalities. $a$ denotes the audio modality and $v$ denotes the visual modality. $LN$ denotes layer normalization and $MLP$ denotes the feed forward layer. $p$ denotes the \ac{DOA} likelihood and $MSA$ denotes multi-head self-attention, which captures dependencies within a modality and is calculated as follows:
\begin{equation}
    \operatorname{MSA}\left(\mathbf{z}\right) = [\mathbf{h}_{1}; ... ; \mathbf{h}_{n}]\mathbf{W}_{o}
    \label{eq_msa}
\end{equation}
\begin{equation}
    \mathbf{h}_{i} = \operatorname{softmax}\left(\frac{\mathbf{z}\mathbf{W}_{Q_i}\mathbf{W}_{K_i}^{T}\mathbf{z}^{T}}{\sqrt{d_{k}}}\right)\mathbf{z}\mathbf{W}_{V_i}
    \label{eq_att}
\end{equation}
\noindent where Eq. (\ref{eq_msa}) aggregates the outputs from multiple heads and Eq. (\ref{eq_att}) defines the attention calculation within each head. $n$ is the number of heads. $\mathbf{W}_{Q}, \mathbf{W}_{K}$ and $\mathbf{W}_{V}$ are learnable matrix for calculating query, key and value. $\mathbf{W}_{o}$ is the learnable output matrix and $\operatorname{softmax}$ is used to normalize the attention matrix.

If the speaker is not in the camera view, adding visual information is not beneficial for localization. Thus we split each training batch into two parts. The part where the speaker is in the camera view and the other part where is speaker is out of the view. For the first part ($I_{av}$), both audio and visual modality are useful for localization, which are passed through separate encoders to obtain audio and visual representations. These representations are fused for \ac{DOA} prediction. For the other part ($I_{ao}$), only the audio information is used. The audio features go through the audio encoder and the final classifier to obtain the \ac{DOA} without the multimodal fusion (denoted as the dotted line in Fig. \ref{model}). 
\begin{equation}
    \hat{p} = \operatorname{MLP}\left( \mathbf{z}_{a}^{D}\right)
\end{equation}
This allows the model to leverage both audio and visual cues when available, while relying solely on audio when the speaker is not in view. Thus the redundant visual information will not affect the \ac{DOA} prediction. The detailed process of \ac{DOA} estimation is shown in Algorithm \ref{alg:A}.

\begin{algorithm}[t]
\SetAlgoLined
\caption{Egocentric Audio-Visual DOA Estimation}
\label{alg:A}
\SetKwInOut{Input}{Input}
\SetKwInOut{Output}{Output}
\Input{$a_{1},..., a_{BatchSize}$, $v_{1},..., v_{BatchSize}$, $AudioEncoder$, $VisualEncoder$, $I_{av} = \varnothing$, $I_{ao} = \varnothing$}
\Output{$\hat{p}$}

\For{$i \leftarrow 1$ \KwTo $BatchSize$}{
    \If{speaker is not in the field of view in $a_{i}$}{
        $I_{ao} \leftarrow i$
    }
    \Else{
        $I_{av} \leftarrow i$
    }
}

$\hat{p}_{ao} \leftarrow MLP(AudioEncoder(a_{I_{ao}}))$\;
$\hat{p}_{av} \leftarrow MLP(MSA(AudioEncoder(a_{I_{av}})), \break VisualEncoder(v_{I_{av}}))$\;
\Return $\hat{p} = \hat{p}_{ao} \cup \hat{p}_{av}$\;
\end{algorithm}



\subsection{Learning Objective}
In \cite{wangmicrophone}, each \ac{DOA} is treated as an independent class and \ac{CE} loss is used. However, in \ac{DOA} prediction, each angle is not independent but in order. CE loss ignores the relationship of different angles. Thus, we employ earth mover's distance (EMD) \cite{rubner2000earth} loss, which is used in speech quality assessment \cite{yu2021metricnet}. Recently in \cite{subramanian2022deep}, it has shown competitive effectiveness in sound source localization, even with the requirement of high localizing resolution. Compared to CE loss, EMD loss $\mathcal{L}_{\text{EMD}}$ maintains the consistency of adjacent angles and measures the discrepancy between two distributions, which is calculated as the squared error between the ground truth \ac{DOA} distribution $p$ and the predicted \ac{DOA} distribution $\hat{p}$. $p$ is a Gaussian distribution centered on the ground truth \ac{DOA} $\theta$ with a predefined variance $\sigma$. 
\begin{equation}
    p = \mathcal{N}(\theta, \sigma^2)
\end{equation}
\begin{equation}
    \mathcal{L}_{\text{EMD}} = \sum_{i = 1}^{N} (p_{i} - \hat{p}_{i})^{2} 
\end{equation}
\noindent where $N$ represents the number of \ac{DOA} intervals. In our experiments, we set the resolution of the angle as $1^{\circ}$ thus $N = 360$. 
\subsection{Features}
\subsubsection{Audio Feature Extraction}
For audio feature, we calculate the GCC-PHAT \cite{knapp1976generalized}, which can be used to estimate TDOA between microphones within an array. It is widely used in sound source localization due to its simplicity and effectiveness. Compared to GCC, GCC-PHAT is normalized by magnitude and maintains the phase information, which is more robust against room reverberation and noise. GCC-PHAT is calculated by computing the cross-correlation between microphone signals in the frequency domain and normalizing by the magnitudes:
\begin{equation}
  \mathbf{a}(t, \tau)=\int_{-\infty}^{+\infty} \frac{STFT_{1}(t, f) STFT_{2}^{*}(t, f)}{\left|STFT_{1}(t, f)\right|\left|STFT_{2}^{*}(t, f)\right|} e^{j 2 \pi f \tau} d f
  \label{eq3}
\end{equation}

\noindent where $STFT_{1}$ and $STFT_{2}$ represent the short-term Fourier transform of the audio clips of the paired microphones $(1, 2)$ in a microphone array. $f$ is frequency, $\tau$ is the inter-microphone time lag, $*$ is the complex conjugate. $\mathbf{a} \in \mathbb{R}^{L_{a} \times Z} $ with $L_{a}$ denotes the time length and $Z$ denoting the number of coefficients of time lags. GCC-PHAT is input as a sequence along the first dimension to the audio encoder.

\subsubsection{Visual Feature Extraction}
Previous work \cite{qian2022deep} uses face detectors which detect face bounding boxes or Siamese network \cite{li2022multi} which indicates the most possible speaker position by measuring the similarity between the reference image and test image. However, these methods need to employ an external network for detection, which is computationally expensive.
To lower the computational cost and make the speaker localization end-to-end, we use the image patch embedding as similar in \cite{dosovitskiy2020image}. Specifically, the image $\mathbf{p} \in \mathbb{R}^{3 \times H \times W}$ is split into a sequence of flattened patches $\mathbf{v} \in \mathbb{R}^{L_{v} \times h}$ with convolution operator and patch resolution $r$, where $H$ and $W$ are the width and the height of the image, respectively, $L_{v} = \frac{H \times W}{r^2}$ is the number of image patches and $h$ is the hidden dimension.



\section{Dataset}
\label{Dataset}
\subsection{Difference with Previous Dataset}
There are several datasets used in audio-visual speaker localization and tracking such as AV16.3 \cite{lathoud2004av16}, CAV3D \cite{qian2019multi}, AVDIAR \cite{gebru2016algorithms}, SSLR \cite{he2018deep} and TragicTalkers \cite{berghi2021visually}. 
\begin{enumerate}[i.]
\item AV16.3\cite{lathoud2004av16}: AV16.3 \cite{lathoud2004av16} is recorded with three cameras with a sampling rate of 25 fps and two 8-microphone small-sized circular arrays with a sampling rate of 16 kHz, which is mostly used in audio-visual speaker tracking.
\item CAV3D \cite{qian2019multi}: CAV3D is recorded by Co-located Audio-Visual sensors. It is recorded with an eight-microphone circular array with a sample rate of 96kHz and a camera with 25 fps. Compared to the AV16.3 dataset, scenarios in the CAV3D dataset are more challenging, as it has strong reverberation, contains more frames where speakers are occluded and speakers are outside the \ac{FOV}. 
\item AVDIAR \cite{gebru2016algorithms}: AVDIAR is recorded with a dummy head containing two cameras and six microphones. It is used for speaker diarization and tracking.
\item SSLR \cite{he2018deep}: SSLR is recorded by a robot that contains four microphones. As it lacks a visual modality, previous work \cite{qian2021multi} simulates visual features using 3D position ground truth with a camera projection model.
\item TragicTalkers \cite{berghi2021visually}: This dataset was recorded through a camera rig containing 22 cameras and 38 microphones. There are two speakers in a studio and each speaker takes turns talking. It is used in audio-visual active speaker detection and localization.

\end{enumerate}
However, they are restricted to size or lack the visual modality. For example, in the AV16.3 dataset, the length of the annotated sequence is less than 2 hours, which is not suitable for training a deep learning model. In the SSLR dataset, there are no visual images. In addition, neither of these datasets is in an egocentric view but with a fixed view. So we created a new simulated dataset for Egocentric audio visual speaker localization (Ego-AVSL).
The comparision between our dataset and others is shown in TABLE \ref{t}. 

\subsection{Dataset Generation}
\begin{figure}[tbp]
    \centering
    \includegraphics[width=1\columnwidth]{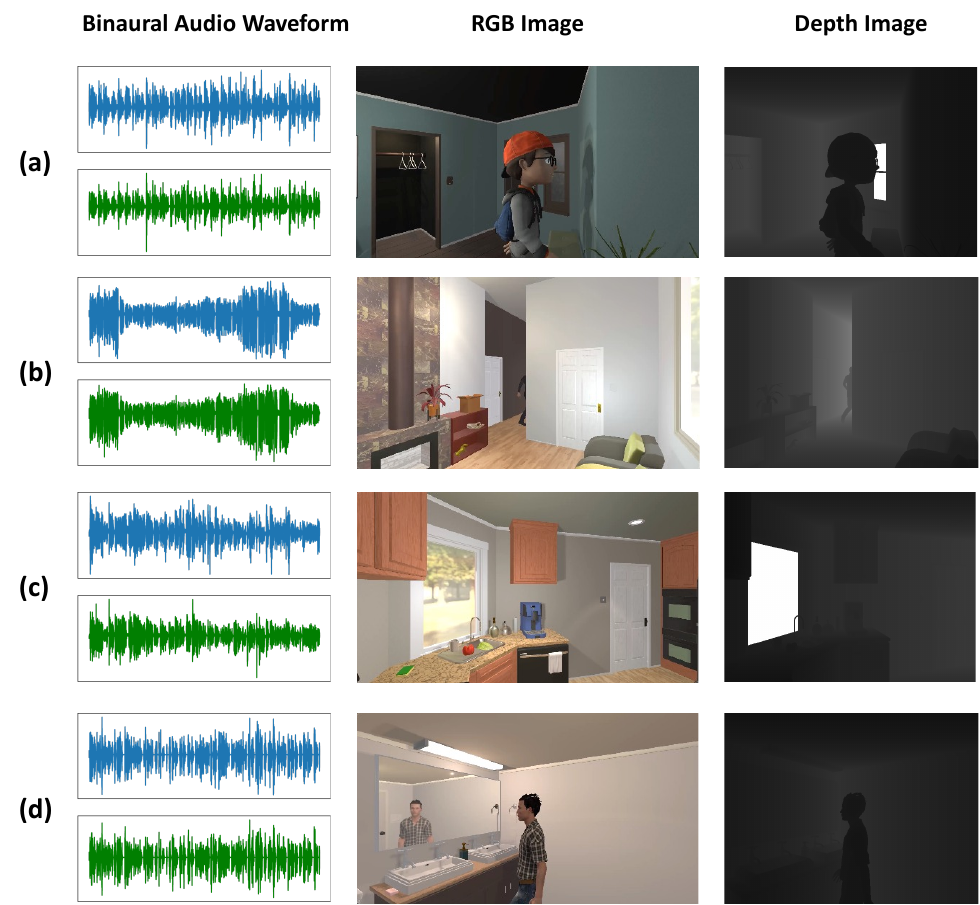}
    \caption{Some scenes in our proposed Ego-AVSL dataset. Different rows show different scenes and different columns show different modalities (Binaural audio waveform, RGB image and depth image). The first row shows a standard scene where the speaker is in the \ac{FOV}. The second row shows an occlusion phenomenon. The third row shows that the speaker is completely out of \ac{FOV}. The fourth row shows that the mirror image of the speaker appears. These scenes show the necessity of using both audio and visual modalities.
    }
    \label{dataset}
\end{figure}

\begin{figure}

  \centering
    \includegraphics[width=0.9\columnwidth]{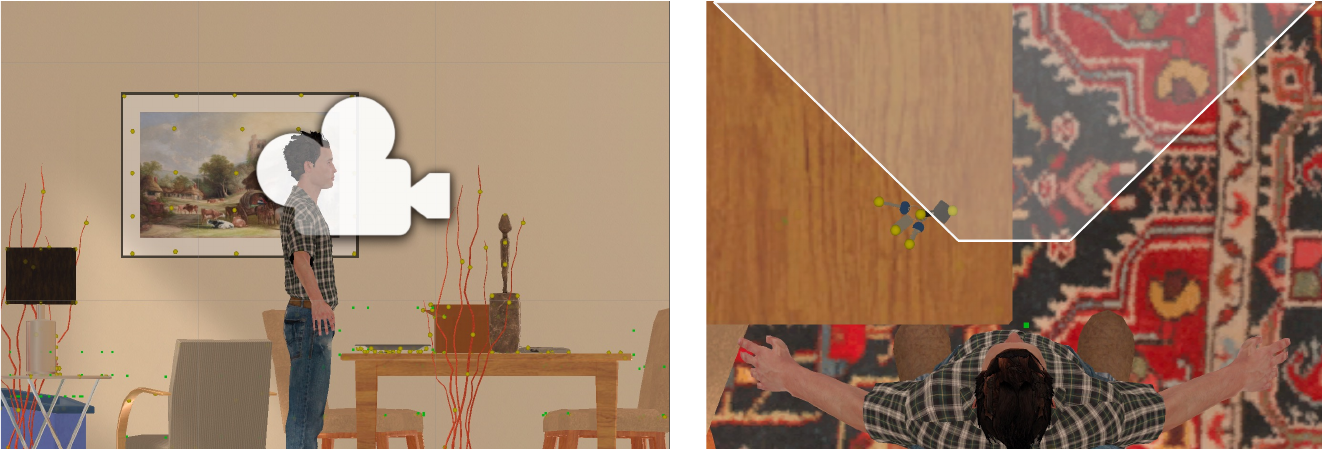}
    \caption{Left: The RGB camera, depth camera and audio receiver are colocated in the front of the wearer's head. Right: The FoV of the RGB camera and the depth camera is $60\degree$. The transparent white area denotes the \ac{FOV}. Thus if the speaker is in the \ac{DOA} of $60 \degree \sim 120 \degree$, the speaker is in the \ac{FOV}.
    }
    \label{camera}
\end{figure}

Our dataset is developed based on Unity\footnote{https://unity.com/} augmented by Resonance Audio\footnote{https://resonance-audio.github.io/resonance-audio/} to generate stereo audio. Resonance Audio can act in the same way as humans hear the sound and simulate head-related transfer function (HRTF), sound occlusions, and reflections, which is used in AR and VR. The characters and motion models are from Mixamo\footnote{https://www.mixamo.com/}. The scenes are from AI2THOR \cite{kolve2017ai2} and the speech clips for the speaker are from Librispeech \cite{panayotov2015librispeech}. For implementation, the speaker is augmented by \textbf{ResonanceAudioSource}, which takes single-channel audio as input and generates spatial binaural audio, and thus the wearer can receive stereo audio. For each scene, the reverberation time is different. The speaker and the camera wearer are initialized with different velocities, orientations and walking periods. In a fixed time interval, both the wearer and the speaker have the possibility to change the orientation. For the label, both the speaker and the wearer have the 3D position and orientation annotation. The speaker also has a 2D bounding box annotation calculated by the Unity camera projection model from the 3D position. The wearer also wears a depth camera, which is in the same position as the RGB camera to capture depth images. The audio receiver is equipped in the camera so that the devices of the three modalities are co-located as shown in Fig. \ref{camera}. 
Both the RGB camera and the depth camera have the \ac{FOV} of 60$\degree$. In this paper, we do not explore the depth modality for estimating the \ac{DOA}, which is left in future work. The simulated dataset includes 330 sequences with a total length of more than 10 hours. The images are captured at 50 fps with a resolution of 1920 $\times$ 1080 while the audio sampling rate is 48k.
Fig. \ref{dataset} illustrates the key scenes with the three involved modalities.

\subsection{Settings of Dataset Simulation}

Both the speaker and the wearer have an initial velocity of $0.5 \sim 1.5 
m/s$. After every moving period ranging from 2 to 4 seconds, they have the possibility of $0.5$ to stand still, or start to walk towards a new direction.



\subsection{Statistic Analysis of Dataset}
\begin{figure}[bp]
    \centering
    \includegraphics[width=\columnwidth]{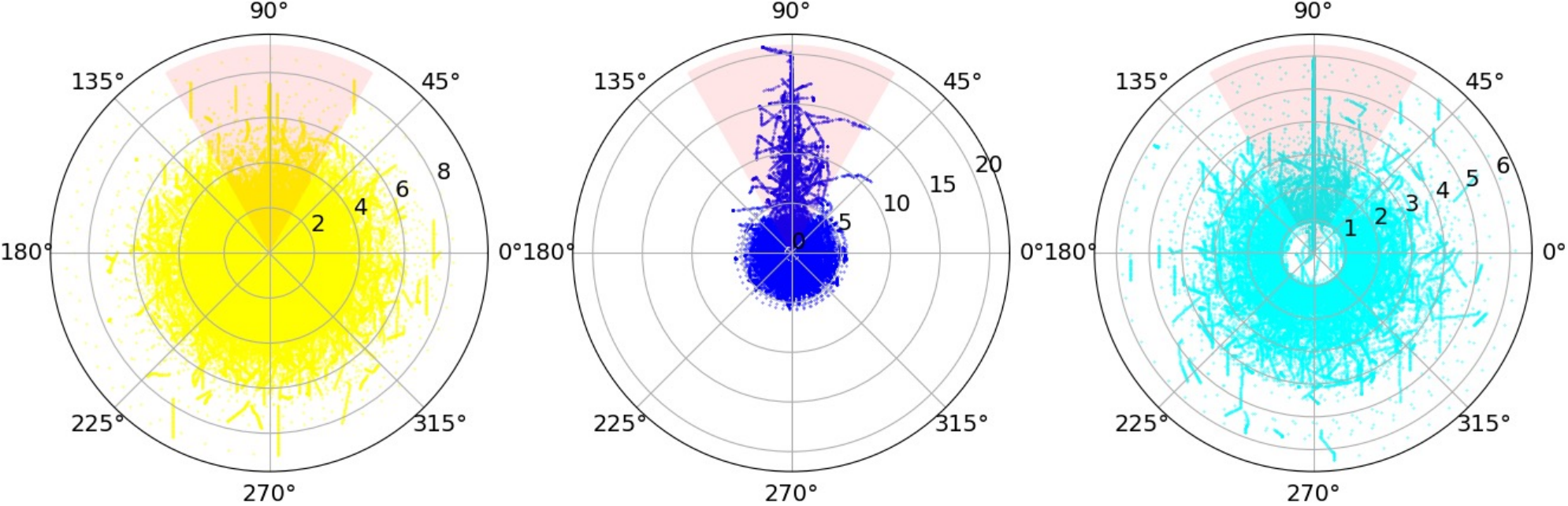}
    \caption{The \ac{DOA} ground truth distribution of the training dataset (yellow), development set (blue) and test set (cyan). The red area denotes the \ac{FOV}.}
    \label{statistic}
\end{figure}

We show the ground truth distribution of \ac{DOA} and distance between the speaker and the wearer in Fig. \ref{statistic}. In most cases, the distance is within 7 meters. However, in the development set, when the speaker is in the \ac{FOV}, there are some cases where the speaker is far away from the wearer by over 15 meters. Besides, the percentage of data where the speaker is in \ac{FOV} is $58.84\%$ and the percentage of data where the speaker is out of \ac{FOV} is $41.16\%$. The simulation follows the realistic scenarios for egocentric AVSL with balanced distribution in different subsets.

\begin{table}[tbp]
\caption{Comparisons between Ego-AVSL and other audio-visual datasets. \textbf{No. Cam} denotes the number of cameras. \textbf{No. Mic} denotes the number of microphones. \textbf{No. Spk} denotes the number of speakers. \textbf{Anno.} denotes the length of the annotated sequence. \textbf{Ego} denotes whether the dataset is in an Ego-centric scenario.}
\centering
\setlength{\tabcolsep}{1.6mm}{
\begin{tabular}{lccccc}
\toprule[1pt]
\specialrule{0em}{1pt}{1pt}

\textbf{Dataset} & \multicolumn{1}{l}{\textbf{No. Cam}} & \multicolumn{1}{l}{\textbf{No. Mic}} & \multicolumn{1}{l}{\textbf{No. Spk}} & \multicolumn{1}{l}{\textbf{Anno.}} & \multicolumn{1}{l}{\textbf{Ego}} \\ \hline\specialrule{0em}{1pt}{1pt}
AV16.3           & 3                                    & 16                                   & 1-3                                  & \textless{}2h                      &  $\times$                               \\ \hline\specialrule{0em}{1pt}{1pt}
CAV3D            & 5                                    & 8                                    & 1-3                                  & \textless{}2h                      &  $\times$                                \\ \hline\specialrule{0em}{1pt}{1pt}
AVDIAR           & 2                                    & 6                                    & 1-4                                  & \textless{}0.5h                    &  $\times$                                \\ \hline\specialrule{0em}{1pt}{1pt}
TragicTalkers      & 22                                   & 38                                   & 1                                    & $\sim$2.5h                         & $\times$                                 \\ \hline\specialrule{0em}{1pt}{1pt}
Ego-AVSL         & 1                                    & 2                                    & 1                                    & $\sim$10h                          & $\checkmark$                                 \\ \bottomrule[1pt]
\end{tabular}}
\label{t}
\end{table}

\section{Experiments}
\label{experiments}
\subsection{Implementation Details and Evaluation Metrics}

\subsubsection{Model Configuration and Training Details}
We use 265 sequences for training, 30 sequences for validation, and 20 sequences for testing. We divide each sequence into chunks. Each chunk has one image frame, one audio clip, and one depth image. We extract one chunk for every two image frames to mitigate data repetition. The length of the audio clip is equal to 25 image frames. After splitting, the number of chunks for training, validation and testing is around 460,000, 130,000 and 90,000, respectively. For calculating GCC-PHAT \cite{knapp1976generalized} of the audio clip, the length of fast Fourier transfer window is 1024, the hop size is 320 and the number of coefficients of time lags is 96. For visual features, the frames are resized to 224 $\times$ 224. Each frame is split into 16 $\times$ 16 patches. For the transformer encoder, we choose 2 layers, 4 attention heads, 256 intermediate dimensions, and 128 hidden dimensions. For the training process, the batch size is 512 and the learning rate is 1e-3. The epoch is 30 with the early stop mechanism.

\subsubsection{Evaluation Metrics}
We use Absolute Error (AE) and Accuracy to measure the performance of models.  AE is calculated as follows:
\begin{equation}
    \mathrm{AE}=|\theta-\hat{\theta}|
\end{equation}

Where $\theta$ is the ground truth degree and $\hat{\theta}$ is the predicted degree, selected as the maximum of $\hat{p}$. Since the direction of arrival value is cyclic. $AE$ should range from 0 to 180$\degree$. So if $|\theta-\hat{\theta}|$ is larger than 180$\degree$, we use $360\degree - |\theta-\hat{\theta}|$ as $AE$.
Accuracy is calculated as the percentage of the correct predictions. The prediction is regarded as correct if the AE between the prediction and the ground truth \ac{DOA} is less than two degrees. 

\subsection{Experimental Results}
\subsubsection{Baselines}
We reimplement and compare the following baseline methods:

\begin{enumerate}[i.]
\item SRP-PHAT \cite{brandstein1997robust}: Steered Response Power Phase Transform, also named Global Coherence Field. It is a signal processing-based method which adds up the GCC-PHAT from multiple microphone pairs within one microphone array and locates the sounding object with the maximum response power.
\item MLPGCC \cite{he2018deep}: A learning-based method with fully-connected layers which takes flatted and concatenated GCC-PHAT from all possible microphone pair combinations as input. 
\item MLPAVC \cite{qian2021multi}: A learning-based method with fully connected layers which takes the early concatenation of GCC-PHAT and visual features as input.
\item MLPAVAM \cite{qian2021multi}: Based on MLPAVC, a dynamic weighting mechanism is introduced to determine the importance of the audio and visual modality. 
\item Multimodal Transformer (Multimodal Trn): Transformer with the input of the combination of $CLS$, the GCC-PHAT and the split image patches. The output corresponding to the $CLS$ is passed to \ac{DOA} classifier for \ac{DOA} prediction.
\end{enumerate}
\begin{table}[tbp]
\caption{Experimental Results (No. Params denotes the number of trainable model parameters with the unit MegaBytes. `-' denotes information not applicable.)}
\centering
\setlength{\tabcolsep}{0.4mm}{
\begin{tabular}{llllll}
\toprule[1pt]
 & \multicolumn{2}{c}{\textbf{Ego-AVST}}                            & \multicolumn{2}{c}{\textbf{Ego-AVST-Noise}} & \textbf{Model Size}   \\ \hline
\multicolumn{1}{l}{} & \multicolumn{1}{l|}{\textbf{Accuracy}$(\%)$} & \multicolumn{1}{l|}{\textbf{AE}($\degree$)} & \multicolumn{1}{l|}{\textbf{Accuracy}$(\%)$} & \textbf{AE}($\degree$) & \multicolumn{1}{|l}{\textbf{No. Params}}\\ \hline\specialrule{0em}{1pt}{1pt}
SRP-PHAT\cite{brandstein1997robust}                &  \multicolumn{1}{c}{10.060}                             & \multicolumn{1}{c}{142.00}                      & \multicolumn{1}{c}{10.850}                              & \multicolumn{1}{c}{141.35} & \multicolumn{1}{c}{-}     \\ \hline\specialrule{0em}{1pt}{1pt} 
MLPGCC \cite{he2018deep}                &  \multicolumn{1}{c}{77.519}                             & \multicolumn{1}{c}{14.237}                      & \multicolumn{1}{c}{74.281}                              & \multicolumn{1}{c}{21.478} & \multicolumn{1}{c}{2.4M}   \\ \bottomrule[1pt]\specialrule{0em}{1pt}{1pt}
MLPAVC \cite{qian2021multi}               &  \multicolumn{1}{c}{82.788}                                & \multicolumn{1}{c}{10.530}                         &  \multicolumn{1}{c}{\textbf{80.729}}                            &  \multicolumn{1}{c}{16.661} &  \multicolumn{1}{c}{2.5M}   \\ \hline\specialrule{0em}{1pt}{1pt}
MLPAVAM \cite{qian2021multi}              &  \multicolumn{1}{c}{82.992}                             &  \multicolumn{1}{c}{10.387}                        & \multicolumn{1}{c}{80.540}                             & \multicolumn{1}{c}{16.265}  & \multicolumn{1}{c}{2.6M}   \\ \hline\specialrule{0em}{1pt}{1pt}

Multimodal Trn               &  \multicolumn{1}{c}{86.591}                             &  \multicolumn{1}{c}{4.157}                        & \multicolumn{1}{c}{77.135}                              & \multicolumn{1}{c}{17.244}  & \multicolumn{1}{c}{3.0M}   \\ \hline\specialrule{0em}{1pt}{1pt}
Ours                  & \multicolumn{1}{c}{\textbf{89.951}}                              & \multicolumn{1}{c}{\textbf{3.859}}                         &  \multicolumn{1}{c}{77.853}                             & \multicolumn{1}{c}{\textbf{15.044}}  & \multicolumn{1}{c}{1.8M}   \\ \bottomrule[1pt]

\end{tabular}}
\label{t2}
\end{table}
\subsubsection{Analysis}
\label{analysis}
We show the localization results in TABLE \ref{t2}. The first three methods are audio methods while the remaining are audio-visual methods. In addition to the results on our proposed Ego-AVST, we also report the results on the noisy version of the test set of Ego-AVST. The noisy version is created by adding noise from DEMAND \cite{thiemann2013demand} which is recorded in real scenarios such as offices, subways and meeting rooms with an SNR level of 20 dB. Compared to the learning-based methods, the traditional SRP-PHAT method fails and the learning-based methods are more robust in the ego-centric scenario. MLPAVAM and MLPAVC outperform MLPGCC, showing the advantages of using both audio and visual modalities as the visual modality can help to localize the speaker. It can be seen that our proposed method outperforms the learning-based three baselines to a large extent, showing the modeling capacity of Transformer. The multimodal transformer takes the concatenation of GCC-PHAT and image patches as input. It does not show a good result, indicating that it is not suitable to encode the audio and visual features in one Transformer as there exists a domain gap between the two modalities. 

\begin{figure}[tbp]
    \centering
    \includegraphics[width=0.7\columnwidth]{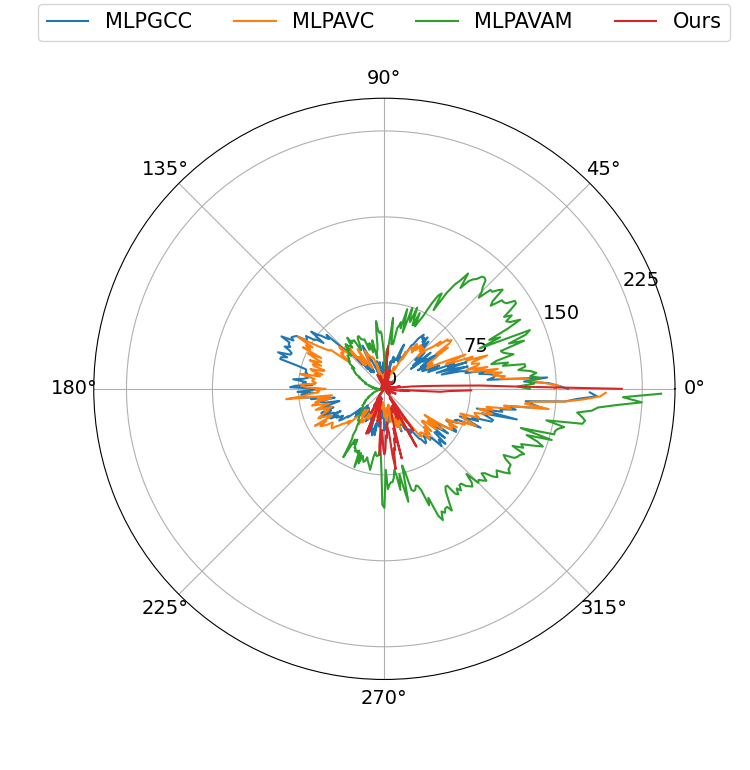}
    \caption{The \ac{DOA} error distribution of our method and baselines.}
    \label{doaerror}
\end{figure}

We also show the \ac{DOA} error distribution in Fig. \ref{doaerror}, which demonstrates the error over all \ac{DOA} in the test set. We can see that our method shows lower errors over most \ac{DOA} except in the range of 260 $\sim$ 300$\degree$. Besides, our method shows larger errors in the range of 180 $\sim$ 360$\degree$ than those in the range of 60 $\sim$ 120$\degree$. The possible reason is that when the target \ac{DOA} is in the range of 60 $\sim$ 120$\degree$, the visual information is available and beneficial for the localization. And the audio and visual modality can work jointly to strengthen the model performance. When the target \ac{DOA} is in the range of 180 $\sim$ 360$\degree$, only audio is used and the localization resolution of audio is low. 

\begin{figure}[tbp]
    \centering
    \includegraphics[width=0.95\columnwidth]{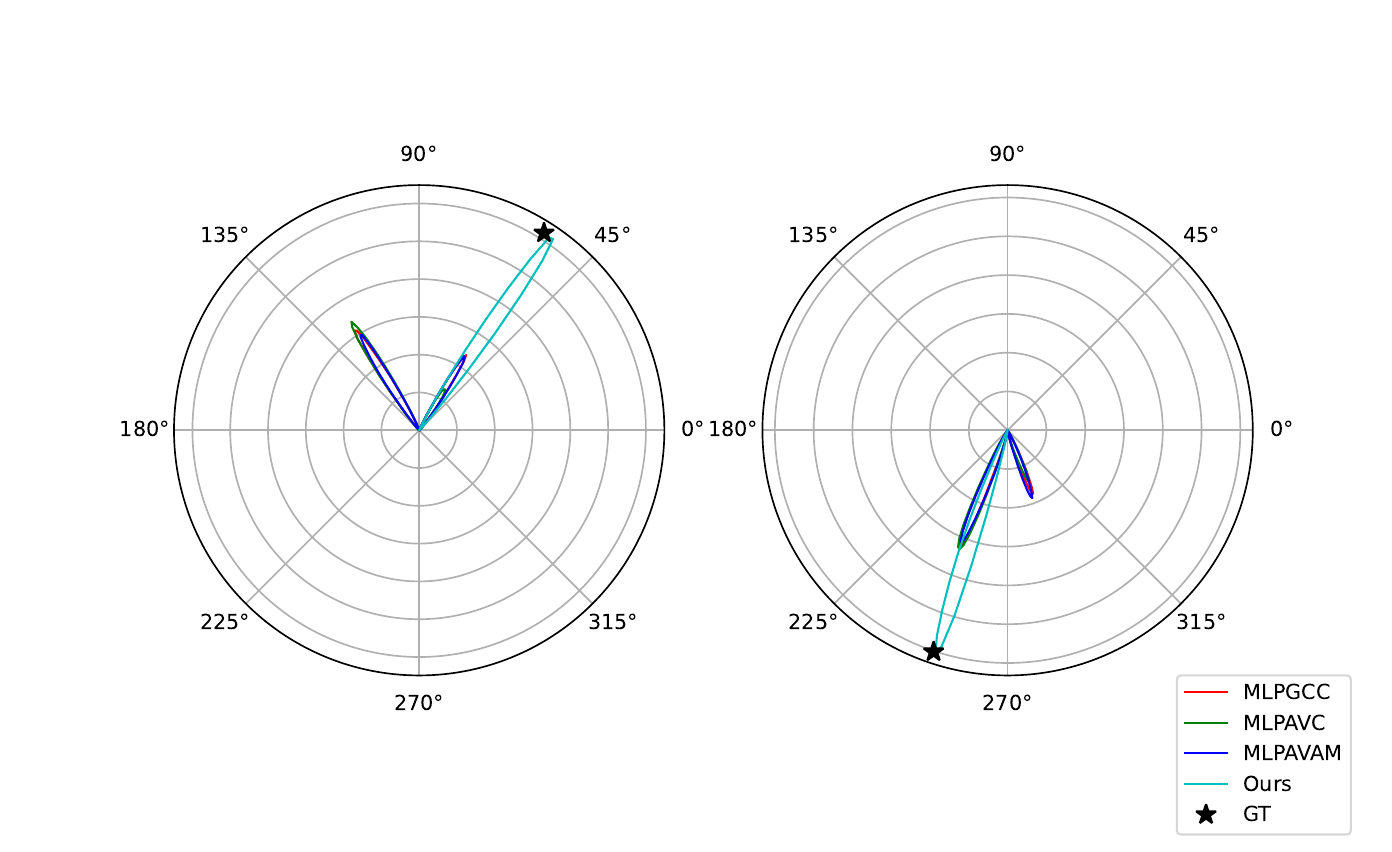}
    \caption{The \ac{DOA} estimation of two examples. The left figure shows the case where the speaker is in the front of the wearer while the right figure shows the case where the speaker is in the back. }
    \label{doaestimation}
\end{figure}

We show two examples of \ac{DOA} estimation by the baseline methods and our method in Fig. \ref{doaestimation}. The angle corresponding to the largest value within one color indicates the \ac{DOA} estimation. In the first figure, the estimations of baselines deviate a lot from the ground truth. And in the second figure, the estimation of our method is closer to the ground truth compared to the baseline estimation. However, from TABLE \ref{t2}, it can be seen that the MLP-based methods show better robustness against background noise than the Transformer-based methods.

\subsection{Ablation Study}
We do the ablation study to show the helpfulness of the EMD loss and the audio-visual fusion. The results of performance variation are listed in TABLE \ref{t3}. It is demonstrated that the EMD loss is superior to the Cross Entropy loss as the latter cannot model the relationship between different \ac{DOA}. Without audio modality, the model performance degrades significantly, as the speaker is often out of the camera's \ac{FOV} and the visual modality is of no use for localization. Compared to the model with audio only input, our model shows lower localization errors as the visual modality helps to localize the speaker when the speaker is in the view.

We also show the effectiveness of separate training. The second method \textbf{w/o Separate} means we directly input both the audio and visual stream into the model without recognizing the audio only part and audio-visual part. The results turn out that the performance is not as good as our method, as the visual information is not beneficial for localization when the speaker is out of \ac{FOV}.

We show the effects of visual modality in TABLE \ref{t4}. It can be seen that with visual information, the model performance degrades a little in the scenario where the speaker is out of the \ac{FOV}. Even though the visual data does not take part in the training process when the speaker is out of the view, it influences the model optimization in the training batches where the speaker is in the view. However, with the help of visual modality, the model performance improves in the scenario where the speaker is in the camera view, especially lowering the MAE from 6.830 to 3.727. 

In addition, we compare the multi-modal fusion methods. In TABLE \ref{t3}, \textbf{AV Trn $\times$} denotes the audio visual transformer with multiplication fusion. The posterior from the audio encoder and visual encoder are multiplied together to obtain the final posterior. The performance of \textbf{AV Trn $\times$} is not good than that of our method. The possible reason is that if the posterior of one modality is not reliable it will have a negative impact on the final decision of the model. However, the cross modality attention module used in our model will assign a hidden weight to each modality based on the modality interaction. If one modality is not reliable, a lower weight will be assigned, which will mitigate the negative impact.

\begin{table}[tbp]
\caption{Ablation Study (\textbf{w/o} denotes `without', \textbf{Trn} denotes `Transformer' and `$\times$' denotes multiplication of multimodal likelihood.)}
\centering
\setlength{\tabcolsep}{0.7mm}{
\begin{tabular}{lllll}
\toprule[1pt]
 & \multicolumn{2}{c}{\textbf{Ego-AVST}}                            & \multicolumn{2}{c}{\textbf{Ego-AVST-Noise}}  \\ \hline
\multicolumn{1}{l}{} & \multicolumn{1}{l|}{\textbf{Accuracy}$(\%)$} & \multicolumn{1}{l|}{\textbf{AE}($\degree$)} & \multicolumn{1}{l|}{\textbf{Accuracy}$(\%)$} & \textbf{AE}($\degree$) \\ \hline\specialrule{0em}{1pt}{1pt}
w/o EMD               &  \multicolumn{1}{c}{50.671}                             & \multicolumn{1}{c}{56.644}                      & \multicolumn{1}{c}{50.671}                              & \multicolumn{1}{c}{56.644}    \\ \hline\specialrule{0em}{1pt}{1pt}
w/o Separate               &  \multicolumn{1}{c}{86.446}                                & \multicolumn{1}{c}{4.411}                         &  \multicolumn{1}{c}{77.288}                             &  \multicolumn{1}{c}{16.213}   \\ \hline\specialrule{0em}{1pt}{1pt}
Visual Only Trn               &  \multicolumn{1}{c}{57.989}                                & \multicolumn{1}{c}{48.490}                         &  \multicolumn{1}{c}{57.989}                             &  \multicolumn{1}{c}{48.490}   \\ \hline\specialrule{0em}{1pt}{1pt}
Audio Only Trn               &  \multicolumn{1}{c}{89.762}                                & \multicolumn{1}{c}{6.297}                         &  \multicolumn{1}{c}{77.460}                             &  \multicolumn{1}{c}{15.370}   \\ \hline\specialrule{0em}{1pt}{1pt}
AV Trn $\times$               &  \multicolumn{1}{c}{87.435}                                & \multicolumn{1}{c}{4.480}                         &  \multicolumn{1}{c}{77.089}                             &  \multicolumn{1}{c}{17.177}   \\ \hline\specialrule{0em}{1pt}{1pt}
Ours               &  \multicolumn{1}{c}{\textbf{89.951}}                             &  \multicolumn{1}{c}{\textbf{3.859}}                        & \multicolumn{1}{c}{\textbf{77.853}}                              & \multicolumn{1}{c}{\textbf{15.044}}    \\ \bottomrule[1pt]

\end{tabular}}
\label{t3}
\end{table}

\begin{table}[tbp]
\caption{Effect of the visual signals (\textbf{AO} denotes `Audio Only' and \textbf{AV} denotes `Audio Visual')}
\centering
\setlength{\tabcolsep}{1mm}{
\begin{tabular}{lllllll}
\toprule[1pt]
                      & \multicolumn{2}{c}{\textbf{In the View}}                          & \multicolumn{2}{c}{\textbf{Out of the View}}                      & \multicolumn{2}{c}{\textbf{Overall}}         \\ \hline
\multicolumn{1}{l|}{} & \multicolumn{1}{l|}{\textbf{Acc}$(\%)$} & \multicolumn{1}{l|}{\textbf{AE}($\degree$)} & \multicolumn{1}{l|}{\textbf{Acc}$(\%)$} & \multicolumn{1}{l|}{\textbf{AE}($\degree$)} & \multicolumn{1}{l|}{\textbf{Acc}$(\%)$} & \textbf{AE}($\degree$) \\ \hline\specialrule{0em}{1pt}{1pt}
AO            & \multicolumn{1}{c}{92.625}       &\multicolumn{1}{c}{6.830}  & \multicolumn{1}{c}{\textbf{75.822}}                              & \multicolumn{1}{c}{\textbf{3.689}}                         & \multicolumn{1}{c}{89.762}                              & \multicolumn{1}{c}{6.297}    \\ \hline\specialrule{0em}{1pt}{1pt}
AV          &   \multicolumn{1}{c}{\textbf{93.596}}                             & \multicolumn{1}{c}{\textbf{3.727}}                         &  \multicolumn{1}{c}{72.466}                             & \multicolumn{1}{c}{4.545}                         &  \multicolumn{1}{c}{\textbf{89.951}}                             & \multicolumn{1}{c}{\textbf{3.859}}    \\ \bottomrule[1pt]
\end{tabular}}
\label{t4}
\end{table}

\subsection{Visualization Results}
\begin{figure*}[tbp]
\centering
\includegraphics[scale=0.53]{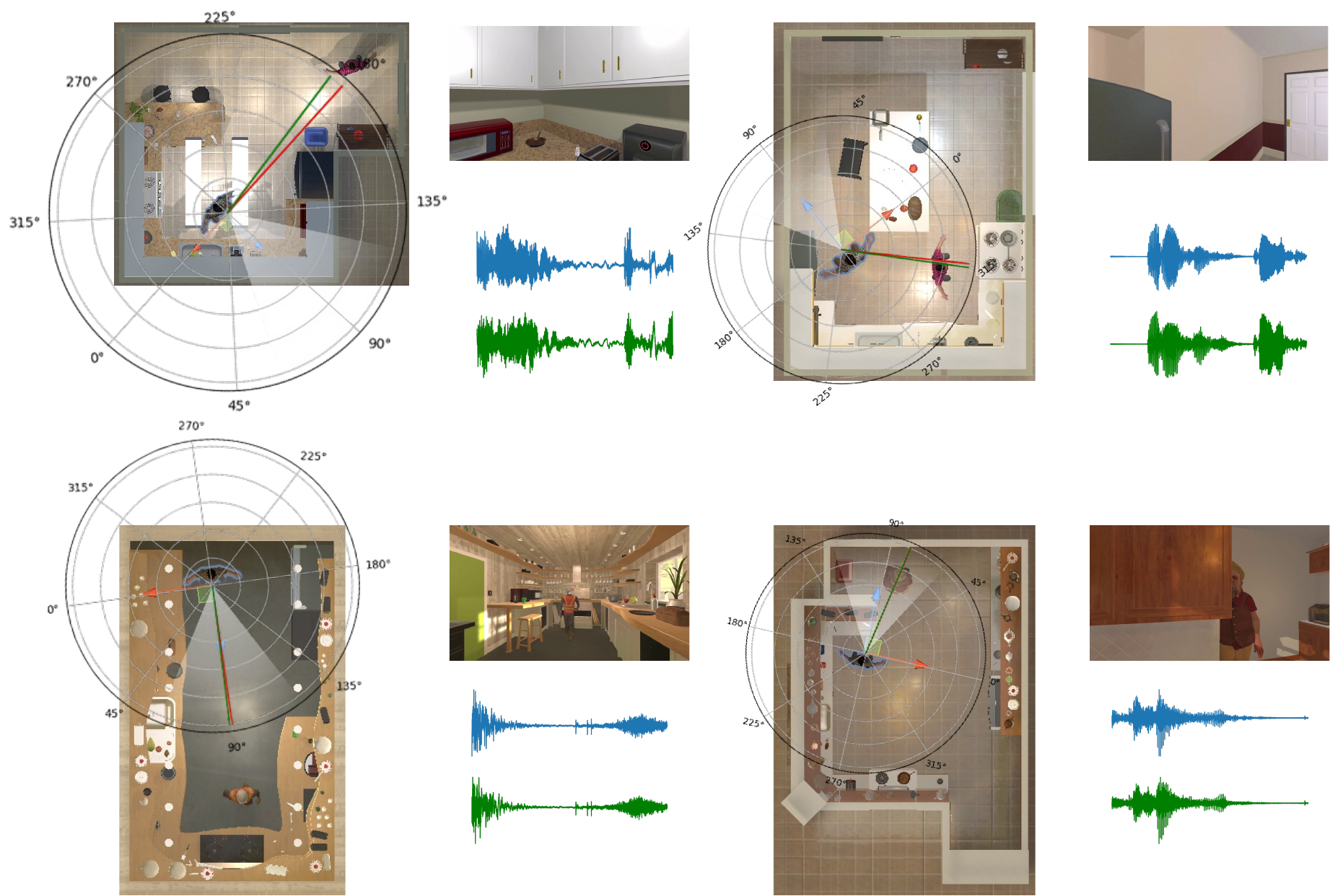}
\caption{Visualization results of \ac{DOA} estimations with the egocentric image and binaural audio. The blue arrow and the red arrow show the positive directions of the north and the east, respectively. The fan areas with white gradient starting from the origin in the polar coordinates show the \ac{FOV}. The green straight line is the estimation results while the red straight line is the ground truth. In the first row, there are two examples where speakers are out of \ac{FOV}. In the second row, the scenarios where the speakers are in the \ac{FOV} are shown.}
\label{visual}
\end{figure*}
In Fig. \ref{visual}, we show some visualization results of \ac{DOA} estimation with audio and visual input stream to show the effectiveness of the proposed model. In the first row, the speakers are out of \ac{FOV} and no speaker appears in the egocentric image. The \ac{DOA} are estimated correctly thanks to the audio modality. In the second row, the speakers are in the \ac{FOV} with partial occlusions in the second egocentric image. Our model can also localize the speaker accurately by leveraging both audio and visual modalities.

\begin{figure*}[tbp]
    \begin{subfigure}[t]{0.33\textwidth}
           \centering
           \includegraphics[width=\textwidth]{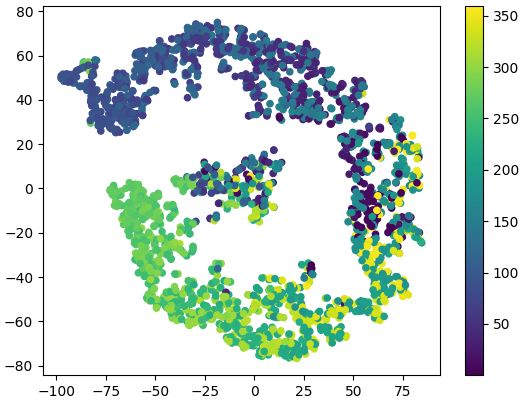}
            \caption{MLPGCC \cite{he2018deep}}
            \label{fig:a}
    \end{subfigure}
    \begin{subfigure}[t]{0.33\textwidth}
            \centering
            \includegraphics[width=\textwidth]{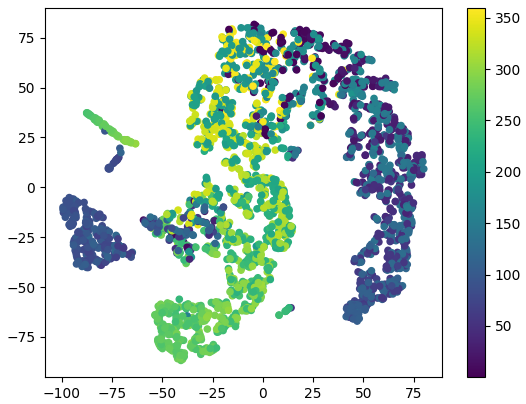}
            \caption{MLPAVAM \cite{qian2021multi}}
            \label{fig:b}
    \end{subfigure}
    \begin{subfigure}[t]{0.33\textwidth}
            \centering
            \includegraphics[width=\textwidth]{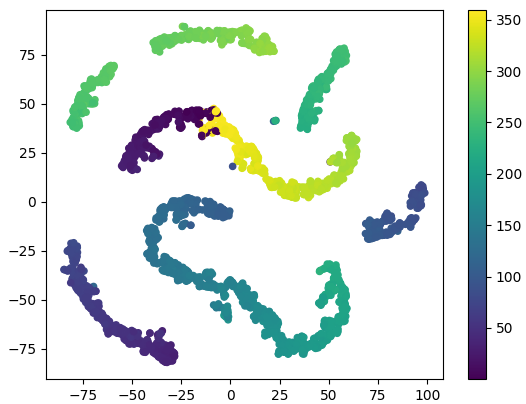}
            \caption{Ours}
            \label{fig:c}
    \end{subfigure}
    \caption{The t-SNE visualization of audio only method MLPGCC\cite{he2018deep}, audio visual method MLPAVAM\cite{qian2021multi} and our method.}
\label{f9}
\end{figure*}

We also show the t-SNE visualization of the extracted features by the baseline methods MLPGCC \cite{he2018deep}, MLPAVAM \cite{qian2021multi} and our method in Fig. \ref{f9}. We use the features before the \ac{DOA} classifier for visualization. As discussed in \cite{qian2022audio}, the \ac{DOA} is consistent between $360\degree$ and $0\degree$, thus the yellow points and dark blue points should be close to each other. It is observed in Fig. \ref{fig:a} and Fig. \ref{fig:b}, there are some points of different colors clustered in the center which are not well classified. While in Fig. \ref{fig:c}, different groups of points denoting different \ac{DOA} ranges are well classified. 

\section{Extension to Multiple Speakers in Real Scenarios}
\label{extension_multiple}
\subsection{Speaker Activity Detection}
\label{extension}
Although the proposed simulated dataset mitigates the problem of manual annotation and privacy issues, there are some limitations of the proposed dataset. On the one hand, there is only one speaker. While in the real scenarios there exists scenarios that multiple speakers talk at the same time. On the other, the simulated dataset has domain gaps with the real dataset and real scenario. Thus we evaluate the model performance on real dataset. 

\begin{table}[tbp]
\caption{Experimental results on EasyCom Dataset for egocentric active speaker localization. Experimental results of \textit{AV(cor)}, \textit{AV(spec)}, \textit{DOA}, \textit{DOA+image} and \textit{AV-rawaudio} are imported from \cite{jiang2022egocentric}. $\dagger$ denotes that the model is trained from scratch. $\ddagger$ denotes that the model is initialized with the weights pretrained on the proposed simulated dataset. }
\centering
\setlength{\tabcolsep}{4mm}{
\begin{tabular}{lrrrr}
\toprule[1pt]
\textbf{Methods}
            & \multicolumn{1}{l}{\textbf{Mean E1}} & \multicolumn{1}{l}{\textbf{Std1}} & \multicolumn{1}{l}{\textbf{Mean E2}} & \multicolumn{1}{l}{\textbf{Std2}} \\ \hline\specialrule{0em}{1pt}{1pt}
AV(cor)     & 16.77                       & 12.63                    & 6.56                        & 8.77                     \\ \hline\specialrule{0em}{1pt}{1pt}
AV(spec)    & 8.81                        & 9.63                     & 6.21                        & 6.89                     \\ \hline\specialrule{0em}{1pt}{1pt}
DOA         & 129.82                      & 18.26                    & 46.45                       & 21.50                    \\ \hline\specialrule{0em}{1pt}{1pt}
DOA+image   & 66.81                       & 7.89                     & 36.48                       & 8.97                     \\ \hline\specialrule{0em}{1pt}{1pt}
AV-rawaudio & 40.14                       & 10.55                    & 140.75                      & 19.58                    \\ \hline\specialrule{0em}{1pt}{1pt}
Ours$^\dagger$       & 9.33               & 12.78                    & 4.72               & 7.15                     \\\hline\specialrule{0em}{1pt}{1pt}
Ours$^\ddagger$         & \textbf{8.00}               & 10.31                    & \textbf{4.49}               & 7.53               \\ \bottomrule[1pt]
\end{tabular}}
\label{t6}
\end{table}

We evaluate the proposed method on the EasyCom dataset \cite{donley2021easycom}. Scenarios in this dataset are introduced in Section \ref{introduction}. Following the settings in \cite{jiang2022egocentric}, we predict a full $360\degree$ spherical speaker localization map with the dimension of $90 * 180$, where $90$ represents the elevation and $180$ represents the azimuth in $2\degree$ resolution. To adapt the proposed model from single speaker localization to multiple speakers localization, we keep most of the model components but the final classification layer. After obtaining the concatenated encoded $CLS$ in Fig. \ref{model}, we use two fully connected layers to project the hidden dimension to $4050$ and then reshape it to $(45, 90)$, respectively. Then the two reshaped tensor are stacked to $(2, 45, 90)$ and upsampled to $(2, 90, 180)$, representing the localization map. Non-maximum suppression is first applied to the map to remove the repeated predictions with radius 5 and threshold 0. The selected predictions are matched with ground truth positions with the Hungary algorithm.

The ground truth is obtained by transferring the annotation of 3D positions and quaternions to the $(90, 180)$ map in which $1$ indicates there exists a speaker. The ground truth is augmented by marking the cells near the speaker positions to $1$. The model is trained with Cross Entropy loss. Following \cite{jiang2022egocentric},  we calculate Mean E1, Std1, Mean E2 and Std2 to evaluate the model performance, where Mean E1 and Std1 are the mean and standard deviation of the distance from prediction to ground truth, while Mean E2 and Std2 are the mean and standard deviation of the distance from the ground truth. The former two metrics consider the impact of false positives and the latter two metrics consider the impact of missing targets. EasyCom dataset has 12 sessions, we use sessions 4-12 for training and sessions 1-3 for testing, which keeps the same setting as \cite{jiang2022egocentric}.

Experimental results are shown in TABLE \ref{t6}, where the baseline models from \cite{jiang2022egocentric} are compared. To show the usefulness of the proposed simulated dataset, we provide two versions of the model. The first one is trained from scratch on EasyCom dataset (Ours$\dagger$). The other is initialized with the weights pretrained on the proposed simulated dataset (Ours$\ddagger$) except the final projection layers. We can see that the model initialized with the pretrained weight outperforms the model trained from scratch in terms of Mean E1 and Mean E2, indicating that the model indeed learns some knowledge from the simulated dataset. Besides, it can be seen that our proposed model outperforms the baseline methods in term of Mean E1 and Mean E2, proving the model can handle both the simulated single speaker localization and multi-speaker localization in real scenarios. 

\subsection{Wearer Speech Activity Detection}
In EasyCom dataset, the wearer talks from time to time and the annotation not only contains the speech activity of the participants but the wearer's. We use the trained model in Section \ref{extension} for predicting whether the wearer is talking. Specifically, after training the model for active speaker localization, we use the audio encoder in Section \ref{extension} and add an extra fully connected layer behind the output \textit{CLS} token for binary classification. The model is trained with Cross Entropy loss. We calculate mAP and compare the performance with the baseline model in \cite{jiang2022egocentric}. Experimental results are reported in TABLE \ref{t7}. It can be seen that the proposed model outperforms the baselines, which demonstrates that our model has a good capacity for wearer activity prediction as well. Combined with the experimental results in Section \ref{analysis} and \ref{extension}, it is proven that the proposed model is versatile and is able to predict the audio activity for both speakers and the wearer in both simulated and real scenarios.

\begin{table}[tbp]
\caption{Experimental results on EasyCom Dataset for wearer speech activity. Experimental results of \textit{(cor)}, \textit{(cor+eng)}, \textit{(eng)}, \textit{(spec)}, \textit{(cor)-2ch}, \textit{(spec)-2ch}, \textit{Eng(single channel)} and \textit{AV-rawaudio} are imported from \cite{jiang2022egocentric}.}
\centering
\setlength{\tabcolsep}{4mm}{
\begin{tabular}{lc}
\toprule[1pt]
\textbf{Methods}              & \multicolumn{1}{l}{\textbf{Wearer Speech Activity mAP}} \\ \hline\specialrule{0em}{1pt}{1pt}
(cor)                & 90.20                                          \\ \hline\specialrule{0em}{1pt}{1pt}
(cor+eng)            & 90.13                                          \\ \hline\specialrule{0em}{1pt}{1pt}
(eng)                & 88.89                                          \\ \hline\specialrule{0em}{1pt}{1pt}
(spec)               & 91.69                                          \\ \hline\specialrule{0em}{1pt}{1pt}
(cor)-2ch            & 87.66                                          \\ \hline\specialrule{0em}{1pt}{1pt}
(spec)-2ch           & 90.14                                          \\ \hline\specialrule{0em}{1pt}{1pt}
Eng (single channel) & 76.71                                          \\ \hline\specialrule{0em}{1pt}{1pt}
AV-rawaudio          & 87.29                                          \\ \hline\specialrule{0em}{1pt}{1pt}
Ours                 & \textbf{93.70}                                 \\ \bottomrule[1pt]
\end{tabular}}
\label{t7}
\end{table}

\section{Conclusion}
\label{conclusion}
In this paper, we propose an audio-visual fusion architecture for egocentric audio visual speaker tracking. Experimental results demonstrate the effectiveness of the proposed method and the capability for dealing with out-of-view problems. The ablation study shows that both audio and visual modality is helpful for localization. The visual modality is beneficial when the speaker is in the \ac{FOV}. Experiments on EasyCom dataset also show the effectiveness on the real dataset and the usefulness of the proposed dataset.

For future work, we will explore to add the depth feature for the speaker localization. Actually, the ego-centric tracking scenario is very complicated. There are many problems which may occur in real applications, including motion blur, speaker disappearance, occlusions, surrounding noise, bad illumination conditions \cite{jiang2022egocentric} and so on. In this paper we mainly deal with the challenges of speaker disappearance, occlusions and audio noise. We will continue to focus on the remaining problems.

\begin{bibliography}{mybib}
\bibliographystyle{ieeetr}

\begin{thebibliography}{10}

\bibitem{hampapur2005smart}
A.~Hampapur, L.~Brown, J.~Connell, A.~Ekin, N.~Haas, M.~Lu, H.~Merkl, and
  S.~Pankanti, ``Smart video surveillance: exploring the concept of multiscale
  spatiotemporal tracking,'' {\em IEEE signal processing magazine}, vol.~22,
  no.~2, pp.~38--51, 2005.

\bibitem{loizou2007speech}
P.~C. Loizou, {\em Speech enhancement: theory and practice}.
\newblock CRC press, 2007.

\bibitem{ong2022audio}
J.~Ong, B.~T. Vo, S.~Nordholm, B.-N. Vo, D.~Moratuwage, and C.~Shim,
  ``Audio-visual based online multi-source separation,'' {\em IEEE/ACM
  Transactions on Audio, Speech, and Language Processing}, vol.~30,
  pp.~1219--1234, 2022.

\bibitem{redmon2018yolov3}
J.~Redmon and A.~Farhadi, ``Yolov3: An incremental improvement,'' {\em arXiv
  preprint arXiv:1804.02767}, 2018.

\bibitem{vu2015context}
T.-H. Vu, A.~Osokin, and I.~Laptev, ``Context-aware cnns for person head
  detection,'' in {\em Proceedings of the IEEE International Conference on
  Computer Vision}, pp.~2893--2901, 2015.

\bibitem{li2019dsfd}
J.~Li, Y.~Wang, C.~Wang, Y.~Tai, J.~Qian, J.~Yang, C.~Wang, J.~Li, and
  F.~Huang, ``Dsfd: dual shot face detector,'' in {\em Proceedings of the
  IEEE/CVF Conference on Computer Vision and Pattern Recognition},
  pp.~5060--5069, 2019.

\bibitem{birchfield2005spatiograms}
S.~T. Birchfield and S.~Rangarajan, ``Spatiograms versus histograms for
  region-based tracking,'' in {\em 2005 IEEE Computer Society Conference on
  Computer Vision and Pattern Recognition (CVPR'05)}, vol.~2, pp.~1158--1163,
  IEEE, 2005.

\bibitem{qian2019multi}
X.~Qian, A.~Brutti, O.~Lanz, M.~Omologo, and A.~Cavallaro, ``Multi-speaker
  tracking from an audio--visual sensing device,'' {\em IEEE Transactions on
  Multimedia}, vol.~21, no.~10, pp.~2576--2588, 2019.

\bibitem{knapp1976generalized}
C.~Knapp and G.~Carter, ``The generalized correlation method for estimation of
  time delay,'' {\em IEEE transactions on acoustics, speech, and signal
  processing}, vol.~24, no.~4, pp.~320--327, 1976.

\bibitem{brutti2008localization}
A.~Brutti, M.~Omologo, and P.~Svaizer, ``Localization of multiple speakers
  based on a two step acoustic map analysis,'' in {\em 2008 IEEE International
  Conference on Acoustics, Speech and Signal Processing}, pp.~4349--4352, IEEE,
  2008.

\bibitem{d2012person}
E.~D'Arca, N.~M. Robertson, and J.~Hopgood, ``Person tracking via audio and
  video fusion,'' 2012.

\bibitem{zotkin2001multimodal}
D.~Zotkin, R.~Duraiswami, and L.~S. Davis, ``Multimodal 3-d tracking and event
  detection via the particle filter,'' in {\em Proceedings IEEE workshop on
  detection and recognition of events in video}, pp.~20--27, IEEE, 2001.

\bibitem{he2018deep}
W.~He, P.~Motlicek, and J.-M. Odobez, ``Deep neural networks for multiple
  speaker detection and localization,'' in {\em 2018 IEEE International
  Conference on Robotics and Automation (ICRA)}, pp.~74--79, IEEE, 2018.

\bibitem{he2016deep}
K.~He, X.~Zhang, S.~Ren, and J.~Sun, ``Deep residual learning for image
  recognition,'' in {\em Proceedings of the IEEE conference on computer vision
  and pattern recognition}, pp.~770--778, 2016.

\bibitem{wang2021gcc}
J.~Wang, X.~Qian, Z.~Pan, M.~Zhang, and H.~Li, ``{GCC-PHAT} with
  speech-oriented attention for robotic sound source localization,''
  pp.~74--79, 2021.

\bibitem{cao2019polyphonic}
Y.~Cao, Q.~Kong, T.~Iqbal, F.~An, W.~Wang, and M.~D. Plumbley, ``Polyphonic
  sound event detection and localization using a two-stage strategy,'' {\em
  arXiv preprint arXiv:1905.00268}, 2019.

\bibitem{xu2022ava}
E.~Z. Xu, Z.~Song, S.~Tsutsui, C.~Feng, M.~Ye, and M.~Z. Shou, ``Ava-avd:
  Audio-visual speaker diarization in the wild,'' in {\em Proceedings of the
  30th ACM International Conference on Multimedia}, pp.~3838--3847, 2022.

\bibitem{liu2020forecasting}
M.~Liu, S.~Tang, Y.~Li, and J.~M. Rehg, ``Forecasting human-object interaction:
  Joint prediction of motor attention and egocentric activity,'' {\em Computer
  Vision--ECCV 2020}, vol.~12346, pp.~704--721, 2020.

\bibitem{kazakos2019epic}
E.~Kazakos, A.~Nagrani, A.~Zisserman, and D.~Damen, ``Epic-fusion: Audio-visual
  temporal binding for egocentric action recognition,'' in {\em Proceedings of
  the IEEE/CVF International Conference on Computer Vision}, pp.~5492--5501,
  2019.

\bibitem{grauman2022ego4d}
K.~Grauman, A.~Westbury, E.~Byrne, Z.~Chavis, A.~Furnari, R.~Girdhar,
  J.~Hamburger, H.~Jiang, M.~Liu, X.~Liu, {\em et~al.}, ``Ego4d: Around the
  world in 3,000 hours of egocentric video,'' in {\em Proceedings of the
  IEEE/CVF Conference on Computer Vision and Pattern Recognition},
  pp.~18995--19012, 2022.

\bibitem{damen2018scaling}
D.~Damen, H.~Doughty, G.~M. Farinella, S.~Fidler, A.~Furnari, E.~Kazakos,
  D.~Moltisanti, J.~Munro, T.~Perrett, W.~Price, {\em et~al.}, ``Scaling
  egocentric vision: The epic-kitchens dataset,'' in {\em Proceedings of the
  European Conference on Computer Vision (ECCV)}, pp.~720--736, 2018.

\bibitem{northcutt2020egocom}
C.~Northcutt, S.~Zha, S.~Lovegrove, and R.~Newcombe, ``Egocom: A multi-person
  multi-modal egocentric communications dataset,'' {\em IEEE Transactions on
  Pattern Analysis and Machine Intelligence}, 2020.

\bibitem{fabbri2021motsynth}
M.~Fabbri, G.~Bras{\'o}, G.~Maugeri, O.~Cetintas, R.~Gasparini, A.~O{\v{s}}ep,
  S.~Calderara, L.~Leal-Taix{\'e}, and R.~Cucchiara, ``Motsynth: How can
  synthetic data help pedestrian detection and tracking?,'' in {\em Proceedings
  of the IEEE/CVF International Conference on Computer Vision},
  pp.~10849--10859, 2021.

\bibitem{donley2021easycom}
J.~Donley, V.~Tourbabin, J.-S. Lee, M.~Broyles, H.~Jiang, J.~Shen, M.~Pantic,
  V.~K. Ithapu, and R.~Mehra, ``Easycom: An augmented reality dataset to
  support algorithms for easy communication in noisy environments,'' {\em arXiv
  preprint arXiv:2107.04174}, 2021.

\bibitem{jiang2022egocentric}
H.~Jiang, C.~Murdock, and V.~K. Ithapu, ``Egocentric deep multi-channel
  audio-visual active speaker localization,'' in {\em Proceedings of the
  IEEE/CVF Conference on Computer Vision and Pattern Recognition},
  pp.~10544--10552, 2022.

\bibitem{grumiaux2022survey}
P.-A. Grumiaux, S.~Kiti{\'c}, L.~Girin, and A.~Gu{\'e}rin, ``A survey of sound
  source localization with deep learning methods,'' {\em The Journal of the
  Acoustical Society of America}, vol.~152, no.~1, pp.~107--151, 2022.

\bibitem{vera2018towards}
J.~M. Vera-Diaz, D.~Pizarro, and J.~Macias-Guarasa, ``Towards end-to-end
  acoustic localization using deep learning: From audio signals to source
  position coordinates,'' {\em Sensors}, vol.~18, no.~10, p.~3418, 2018.

\bibitem{adavanne2018sound}
S.~Adavanne, A.~Politis, J.~Nikunen, and T.~Virtanen, ``Sound event
  localization and detection of overlapping sources using convolutional
  recurrent neural networks,'' {\em IEEE Journal of Selected Topics in Signal
  Processing}, vol.~13, no.~1, pp.~34--48, 2018.

\bibitem{michelsanti2021overview}
D.~Michelsanti, Z.-H. Tan, S.-X. Zhang, Y.~Xu, M.~Yu, D.~Yu, and J.~Jensen,
  ``An overview of deep-learning-based audio-visual speech enhancement and
  separation,'' {\em IEEE/ACM Transactions on Audio, Speech, and Language
  Processing}, vol.~29, pp.~1368--1396, 2021.

\bibitem{qian2021multi}
X.~Qian, M.~Madhavi, Z.~Pan, J.~Wang, and H.~Li, ``Multi-target doa estimation
  with an audio-visual fusion mechanism,'' in {\em ICASSP 2021-2021 IEEE
  International Conference on Acoustics, Speech and Signal Processing
  (ICASSP)}, pp.~4280--4284, IEEE, 2021.

\bibitem{wu2023multi}
Y.~Wu, R.~Hu, X.~Wang, and S.~Ke, ``Multi-speaker doa estimation using audio
  and visual modality,'' {\em Neural Processing Letters}, pp.~1--15, 2023.

\bibitem{afouras2020self}
T.~Afouras, A.~Owens, J.~S. Chung, and A.~Zisserman, ``Self-supervised learning
  of audio-visual objects from video,'' in {\em European Conference on Computer
  Vision}, pp.~208--224, Springer, 2020.

\bibitem{wissing2021data}
J.~Wissing, B.~Boenninghoff, D.~Kolossa, T.~Ochiai, M.~Delcroix, K.~Kinoshita,
  T.~Nakatani, S.~Araki, and C.~Schymura, ``Data fusion for audiovisual speaker
  localization: Extending dynamic stream weights to the spatial domain,'' in
  {\em ICASSP 2021-2021 IEEE International Conference on Acoustics, Speech and
  Signal Processing (ICASSP)}, pp.~4705--4709, IEEE, 2021.

\bibitem{qian2022deep}
X.~Qian, Q.~Zhang, G.~Guan, and W.~Xue, ``Deep audio-visual beamforming for
  speaker localization,'' {\em IEEE Signal Processing Letters}, vol.~29,
  pp.~1132--1136, 2022.

\bibitem{qian2022audio}
X.~Qian, Z.~Wang, J.~Wang, G.~Guan, and H.~Li, ``Audio-visual cross-attention
  network for robotic speaker tracking,'' {\em IEEE/ACM Transactions on Audio,
  Speech, and Language Processing}, 2022.

\bibitem{essaadali2016new}
R.~Essaadali and A.~Kouki, ``A new simple unmanned aerial vehicle doppler
  effect rf reducing technique,'' in {\em MILCOM 2016-2016 IEEE Military
  Communications Conference}, pp.~1179--1183, IEEE, 2016.

\bibitem{jia2020localization}
T.~Jia, K.~Ho, H.~Wang, and X.~Shen, ``Localization of a moving object with
  sensors in motion by time delays and doppler shifts,'' {\em IEEE Transactions
  on Signal Processing}, vol.~68, pp.~5824--5841, 2020.

\bibitem{lathoud2004av16}
G.~Lathoud, J.-M. Odobez, and D.~Gatica-Perez, ``Av16. 3: An audio-visual
  corpus for speaker localization and tracking,'' in {\em International
  Workshop on Machine Learning for Multimodal Interaction}, pp.~182--195,
  Springer, 2004.

\bibitem{wu2019dual}
Y.~Wu, L.~Zhu, Y.~Yan, and Y.~Yang, ``Dual attention matching for audio-visual
  event localization,'' in {\em Proceedings of the IEEE/CVF international
  conference on computer vision}, pp.~6292--6300, 2019.

\bibitem{wangmicrophone}
Z.~Wang, X.~Zhao, H.~Rong, Y.~Tong, and J.~Shi, ``Microphone array-based sound
  source localization using convolutional residual network,''

\bibitem{rubner2000earth}
Y.~Rubner, C.~Tomasi, and L.~J. Guibas, ``The earth mover's distance as a
  metric for image retrieval,'' {\em International journal of computer vision},
  vol.~40, no.~2, pp.~99--121, 2000.

\bibitem{yu2021metricnet}
M.~Yu, C.~Zhang, Y.~Xu, S.~Zhang, and D.~Yu, ``Metricnet: Towards improved
  modeling for non-intrusive speech quality assessment,'' {\em arXiv preprint
  arXiv:2104.01227}, 2021.

\bibitem{subramanian2022deep}
A.~S. Subramanian, C.~Weng, S.~Watanabe, M.~Yu, and D.~Yu, ``Deep learning
  based multi-source localization with source splitting and its effectiveness
  in multi-talker speech recognition,'' {\em Computer Speech \& Language},
  vol.~75, p.~101360, 2022.

\bibitem{li2022multi}
Y.~Li, H.~Liu, and H.~Tang, ``Multi-modal perception attention network with
  self-supervised learning for audio-visual speaker tracking,'' in {\em
  Proceedings of the AAAI Conference on Artificial Intelligence}, vol.~36,
  pp.~1456--1463, 2022.

\bibitem{dosovitskiy2020image}
A.~Dosovitskiy, L.~Beyer, A.~Kolesnikov, D.~Weissenborn, X.~Zhai,
  T.~Unterthiner, M.~Dehghani, M.~Minderer, G.~Heigold, S.~Gelly, {\em et~al.},
  ``An image is worth 16x16 words: Transformers for image recognition at
  scale,'' {\em arXiv preprint arXiv:2010.11929}, 2020.

\bibitem{gebru2016algorithms}
I.~D. Gebru, X.~Alameda-Pineda, F.~Forbes, and R.~Horaud, ``Em algorithms for
  weighted-data clustering with application to audio-visual scene analysis,''
  {\em IEEE transactions on pattern analysis and machine intelligence},
  vol.~38, no.~12, pp.~2402--2415, 2016.

\bibitem{berghi2021visually}
D.~Berghi, A.~Hilton, and P.~J. Jackson, ``Visually supervised speaker
  detection and localization via microphone array,'' in {\em 2021 IEEE 23rd
  International Workshop on Multimedia Signal Processing (MMSP)}, pp.~1--6,
  IEEE, 2021.

\bibitem{kolve2017ai2}
E.~Kolve, R.~Mottaghi, W.~Han, E.~VanderBilt, L.~Weihs, A.~Herrasti, D.~Gordon,
  Y.~Zhu, A.~Gupta, and A.~Farhadi, ``Ai2-thor: An interactive 3d environment
  for visual ai,'' {\em arXiv preprint arXiv:1712.05474}, 2017.

\bibitem{panayotov2015librispeech}
V.~Panayotov, G.~Chen, D.~Povey, and S.~Khudanpur, ``Librispeech: an asr corpus
  based on public domain audio books,'' in {\em 2015 IEEE international
  conference on acoustics, speech and signal processing (ICASSP)},
  pp.~5206--5210, IEEE, 2015.

\bibitem{brandstein1997robust}
M.~S. Brandstein and H.~F. Silverman, ``A robust method for speech signal
  time-delay estimation in reverberant rooms,'' in {\em 1997 IEEE International
  Conference on Acoustics, Speech, and Signal Processing}, vol.~1,
  pp.~375--378, IEEE, 1997.

\bibitem{thiemann2013demand}
J.~Thiemann, N.~Ito, and E.~Vincent, ``Demand: a collection of multi-channel
  recordings of acoustic noise in diverse environments,'' in {\em Proc.
  Meetings Acoust}, pp.~1--6, 2013.

\end{thebibliography}
\end{bibliography}


 





\end{document}